\documentclass[doublecolumn]{aa}

\usepackage[utf8]{inputenc}
\usepackage{graphicx}
\usepackage[english]{babel}
\usepackage{natbib,twoopt}
\usepackage[breaklinks=true]{hyperref} 
\bibpunct{(}{)}{;}{a}{}{,}             
\usepackage{amsmath}
\usepackage{placeins}
\usepackage{mathtools}
\usepackage{amssymb}
\usepackage{txfonts}
\usepackage{color}
\usepackage{rotating}
\usepackage{enumitem}
\usepackage[labelfont=bf]{caption}
\usepackage{multirow}
\usepackage{booktabs}
\usepackage{subcaption}
\usepackage{ragged2e}
\usepackage{threeparttable}
\usepackage{placeins}
\usepackage{url}
\hypersetup{colorlinks=true,citecolor=blue}
\usepackage{footnotebackref}

\newcommand{\tab}[1]{Table\,\ref{#1}}

\begin{document}

\title{Measurements and predictions of H$_2$ pressure-broadening coefficients of CO$_2$ absorption lines for exoplanet atmosphere studies}

\author{
  Faten Hendaoui\inst{1,2}
  \and
  Pascale Chelin\inst{3}
  \and
  Xavier Landsheere\inst{3}
  \and
  Hassen Aroui\inst{2}
  \and
  Ha Tran\inst{1}
}

\institute{
Laboratoire de Météorologie Dynamique, IPSL, Sorbonne Université, École Normale Supérieure, Université PSL, École polytechnique, Institut Polytechnique de Paris, CNRS, Paris, France\\
\email{htran@lmd.ipsl.fr}
\and
Université de Tunis, Laboratoire de Spectroscopie et Dynamique Moléculaire, École Nationale Supérieure d’Ingénieurs de Tunis, 5 Av. Taha Hussein, 1008 Tunis, Tunisia
\and
Univ Paris Est Créteil and Université Paris Cité, CNRS, LISA, F-94010 Créteil, France
}

  \date{Submitted in A\&A}
  \abstract{Accurate and comprehensive H$_2$ pressure-induced broadening data for CO$_2$ infrared lines over a wide temperature range are essential for modeling atmospheric opacity of exoplanets. However, available data are currently limited, some of which are affected by large uncertainties. In this work, H$_2$ induced pressure-broadening and pressure-shift coefficients were determined at room temperature for the entire $\nu_3$ band of CO$_2$ in the 4.3 $\mu$m spectral region using a high-resolution Fourier transform spectrometer. In addition, requantized molecular dynamics simulations of the CO$_2$-H$_2$ system were performed using an accurate intermolecular potential. These simulations provide theoretical predictions of H$_2$-broadening coefficients for CO$_2$ lines over a temperature range of 200--1000 K and for rotational quantum number up to $J=120$. The predicted results show very good agreement with the experimental data, with difference of less than 3\%, well below the precision required for exoplanet atmosphere studies. This work provides the first accurate and comprehensive dataset of H$_2$ broadening coefficients for CO$_2$ lines, suitable for modeling of H$_2$-rich exoplanetary atmospheres.
  }
  

\titlerunning{Measurements and predictions of H2 pressure-broadening coefficients of CO2 lines}
\authorrunning{Hendaoui et al.} 

   \maketitle

\section{Introduction}

Reliable H$_2$ pressure-broadening data for a variety of molecules and under wide temperature conditions are essential for accurately modeling exoplanetary atmospheres \citep{hedges2016effect,fortney2019need,tan_h2_2022}. The accuracy of molecular abundances, or more generally atmospheric properties, derived from measured spectra depends on the precision of molecular cross sections, which relies on the knowledge of molecular spectroscopic data, including line position, line intensity, pressure-induced line broadening, and shift due to collisions between the optically active molecule and the main constituents of the considered atmosphere. Several studies demonstrated that the lack of pressure-broadening data leads to large impacts on opacity modeling and the retrieval of atmospheric properties for planets and exoplanets (e.g., \citealt{hedges2016effect, Gharib-Nezhad_2019, garland2019_arXiv, Wiesenfeld_2025}).

However, reliable pressure-broadening information under exoplanet thermophysical conditions remains scarce due to the wide variety of constituents, and temperature and pressure conditions that must be considered. In the absence of available experimental and theoretical data, air-broadened widths are often used with empirical scaling factors \citep{WILZEWSKI_HITRAN_2016, Tan_JGR_2019, tan_h2_2022} to describe the broadening by the main interfering gases in exoplanetary atmospheres such as H$_2$ and He. However, this approach can cause significant discrepancies when compared to accurate measured or calculated broadening data (see \citealt{HENDAOUI_CMDS_CO2He_2026}, for instance). Furthermore, the temperature range encountered in exoplanet atmospheres extends far beyond that of the Earth's atmosphere. For instance, \citet{fortney2019need} noted that for H$_2$- and He-dominated exoplanet atmospheres, the temperature of interest spans from 70 K to 3000 K. 

Carbon dioxide (CO$_2$) is a key molecule in exoplanetary atmospheres. It is detected in a wide range of planets observed with the James Webb Space Telescope (JWST; e.g., \citealt{CO2_JW_Nature_2023, Balmer_2025}). The prominent 4.3 $\mu$m absorption feature is especially valuable as it lies in a spectral region with minimal noise and cloud-haze interference, making it of great interest for both detection and habitability assessments (see \citealt{Wiesenfeld_2025} and references therein). Exploiting this feature is, however, limited by incomplete opacity models, particularly uncertainties in line-broadening and far-wing parameterizations \citep{niraula_impending_2022}. \citet{Wiesenfeld_2025} showed that a precision on pressure-broadening coefficients of better than 10\% is required for JWST exoplanet studies.

Despite their importance, there are few studies devoted to the measurement or calculation of the H$_2$ pressure-broadening coefficients for CO$_2$ lines. In \citep{PADMANABHAN_JQSRT_2014}, H$_2$ broadening coefficients were measured at room temperature for ten CO$_2$ lines, ranging from P(16) to P(34), in the 6317-6335 cm$^{-1}$ region of the 30012←00001 band, using a tunable diode laser. These coefficients were derived from spectra measured for mixtures of CO$_2$, N$_2$, and H$_2$, with a stated uncertainty of about 6\%. The measured coefficients vary significantly with the rotational quantum number, from 0.095 cm$^{-1}$/atm for the P(32) line to 0.121 cm$^{-1}$/atm for the P(20) line. However, no clear rotational dependence is observed, indicating that the actual uncertainty is probably larger than 6\%. Hanson and Whitty \citep{Hanson_Whitty_2014} measured the H$_2$ broadening coefficient for the P(24) line of CO$_2$ at 4957.08 cm$^{-1}$, using wavelength-scanned spectroscopy. These measurements were taken at various temperatures, from 300 K to 700 K, with an uncertainty claimed to be better than 2\%. Their room temperature value for the P(24) line is 0.112 cm$^{-1}$/atm, which closely matches the value of 0.1128 cm$^{-1}$/atm reported by \citet{PADMANABHAN_JQSRT_2014} for the same line. Hanson and Whitty also deduced a temperature dependence exponent of 0.582 for this P(24) line. The experimental investigation by \citet{Burch_JOSA_1969} suggested that H$_2$ broadening is 1.17 times larger than self-broadening for CO$_2$ lines, with the same rotational dependence assumed for broadening by various collision partners. These limited data have been used in various radiative transfer models to calculate the optical thickness of exoplanet atmospheres (e.g., \citealt{Tremblin_2015, Goyal_ATMO_MNRAS_2017, garland2019_arXiv}). 

More recently, \citet{Wiesenfeld_2025} conducted ab initio calculations of H$_2$ pressure broadening for the P(24) line of CO$_2$ in the ground vibrational state. They found that their calculations agree with the measured values of \citep{PADMANABHAN_JQSRT_2014, Hanson_Whitty_2014} within 7\%. \citet{NIE_2025} measured room temperature H$_2$ broadening coefficients for four CO$_2$ lines in the 20012←00001 band, from R(44) to R(50) using a distributed feedback laser. H$_2$ broadening coefficients of CO$_2$ lines in the HITRAN database \citep{gordon_hitran2020_2021} were simply obtained using a scaling factor of 1.5767 on the air-broadening coefficients. Additionally, a constant temperature dependence exponent of 0.58 was applied to all CO$_2$ transitions in HITRAN. It is evident that more accurate and comprehensive measurements and calculations are necessary to provide more reliable line shape data for this system. 

In order to provide a reliable and comprehensive dataset for H$_2$ broadening coefficients of CO$_2$ lines relevant for exoplanet atmosphere studies, we conducted high-resolution laboratory measurements and first-principle calculations. Specifically, a high-resolution Fourier transform spectrometer was used to record spectra of CO$_2$ highly diluted in H$_2$ in the 4.3 $\mu$m spectral range, at room temperature and pressures ranging from 150 mbar to 1060 mbar. The recorded spectra were then analyzed to extract both the pressure-broadening coefficients and the pressure-induced shifts of 61 CO$_2$ lines, from P(62) to R(60). A detailed uncertainty analysis was performed on the obtained parameters, considering uncertainties from various sources. This study represents the first measurement of these coefficients for an entire ro-vibrational band of CO$_2$. In the second step, we used requantized classical molecular dynamics simulations (rCMDSs) to predict CO$_2$ line broadening by H$_2$ over a broad temperature range, from 200 K to 1000 K, which is highly relevant for opacity modeling in exoplanetary atmospheres. These simulations, based on the use of accurate intermolecular potentials and classical theory of molecular collisions, have previously demonstrated the ability to reproduce self-, air-, and He-broadened CO$_2$ line-shape parameters \citep{NGO_Tran_CMDS_CO2_2025, TRAN_CMDS_CO2N2_2025, HENDAOUI_CMDS_CO2He_2026} with relative errors well below 5\% for rotational quantum numbers up to J=160 and temperatures as high as 3000 K. 

The paper is organized as follows. Section 2 focuses on the experimental determination of the pressure-induced line broadening and shift. Section 2.1 describes the experimental setup, the measurement procedure, and the determination of the instrument line shape. The fitting procedure and uncertainty analysis are detailed in Sect. 2.2. The obtained line-broadening and pressure-shift parameters are presented and compared with available data in Sect. 2.3. The rCMDS first-principle calculations for CO$_2$--H$_2$ are described in Sect.3. Section 3.1 details the calculation methods and procedures. The predicted broadening coefficients are then presented and compared with measured values in Sect. 3.2, while the temperature dependence of line broadening is described in Sect. 3.3. Finally, in Sect. 4 we present our the conclusions of this work. 

\section{Measurements}\label{measurements}

\subsection{Experimental setup and recording procedure} 
The high-resolution Fourier transform spectrometer (FTS) at LISA (Bruker IFS-125 HR) was used to record all spectra. The instrument was equipped with a silicon carbide Globar source, a KBr/Ge beamsplitter, and a liquid nitrogen cooled indium antimonide (InSb) detector. The diameter of the FTS iris aperture was set to 2 mm and the focal length of the collimator was 418 mm. No artificial optical weighting was performed (a boxcar function was selected). A 12.3 cm simple-pass absorption cell was used. The Bruker spectral resolution was 0.01 cm$^{-1}$, corresponding to a maximum optical path difference of 90 cm.

The measurements were carried out as follows. The cell was first filled with a very small quantity of CO$_2$ (measured with a 2-Torr full-scale Baratron gauge, MKS 627D model, with an accuracy of 0.15\% of reading) and then completed with H$_2$ to reach the desired total pressure. The latter was measured with a 1000-Torr full-scale Baratron gauge, MKS 628D model, with a stated accuracy of 0.25\% of reading. Commercial carbon dioxide and hydrogen samples from Air Liquide were used without purification (99.9\% natural CO$_2$; 99.9999\% H$_2$). After approximately 20 minutes, once the mixture was homogeneous, a spectrum was recorded by co-adding 3200 scans over a period of 24 hours. The cell was then pumped out and refilled to reach the next desired total pressure. This procedure was adopted to limit any leakage of ambient air into the cell during measurement. The variation in total pressure during 24-hour acquisition was approximately 0.3 mbar (with a mean leak of 2.12 10$^{-4}$ mbar/min). The average temperature during the recordings was about 293.71 K, with a mean variation of 0.3 K per recording (room temperature was used as a proxy for cell temperature). In total, four spectra of mixtures with total pressure ranging from 150 mbar to 1060 mbar were recorded with spectral range spanning from 2200 cm$^{-1}$ to 2500 cm$^{-1}$. This spectral region covers the entire CO$_2$ $\nu$$_3$ band of interest. The experimental pressures and temperatures of the recorded spectra are summarized in \tab{Table1}.

\begin{table}[h]
	\centering
    \caption{Pressures and temperatures of the recorded spectra.}
	\begin{tabular}{ c l l l}
        \hline  \hline 
        Spectrum number & $P_{\mathrm{CO_2}}$(mbar) &  $P_{\mathrm{tot}}$(mbar) &  T(K)\\
        \hline 
        1 & 0.0803  &  0.0803  &  293.50 \\
        2  &  0.0723  &  150.65  &  293.50 \\
        3  &  0.1460  &  506.89  &  293.80 \\
        4  &  0.2205  &  756.73  &  294.25 \\
        5  &  0.2804  &  1062.04  &  293.30 \\
        \hline 
	\end{tabular}

 \label{Table1}
\end{table}

Two empty-cell spectra were recorded with the same optical conditions after pumping down to a pressure below 10$^{-4}$ mbar. These provided the 100\% transmission baseline by removing remaining traces of H$_2$O and CO$_2$ inside the interferometer, and enabled partial removal of the multiplicative channel due to interferences arising from the parallel faces of the cell windows. For each pressure, the corresponding spectrum was obtained by dividing the spectrum recorded with gas by the empty-cell spectrum. A signal-to-noise ratio up to \textasciitilde1500 (see Figure~\ref{fig:fit_exp}) was achieved for the transmission spectra. 

A pure CO$_2$ spectrum at 0.08 mbar was also recorded to determine the instrument line shape (ILS) using the LINEFIT 14.8 software \citep{Hase:99, Hase_amt-5-603-2012}. The results show small asymmetry on the ILS and a modulation close to 1, demonstrating that the different optics of the instrument were well aligned. We performed several tests to determine the optimal spectral interval and sampling step for the ILS. 

For spectral calibration, we used CO$_2$ lines appearing in the empty-cell spectra due to small residual amounts of CO$_2$ inside the FTS. The positions of eight intense lines, determined from fits using a Gaussian profile, were compared with the corresponding line positions given in the HITRAN database \citep{gordon_hitran2020_2021}. The resulting calibration factor was 1.000000467, the standard deviation of the fitted positions being about 2x10$^{-5}$ cm$^{-1}$. As discussed below, because the contribution of weak lines is fixed in the fits using HITRAN spectroscopic line position, accurate spectral calibration is important. 

It should be noted that, due to the very small quantity of CO$_2$ introduced (see Table~\ref{Table1}), possible partial adsorption of the gas on the cell walls, and minor leaks during gas filling, the CO$_2$ partial pressure in the mixture is not known with high precision. However, since the absolute line intensity is not the focus of this work and because the amount of CO$_2$ is very small, the uncertainty in the CO$_2$ pressure does not affect the measured H$_2$ pressure-induced line shape of the considered CO$_2$ transitions. 

\subsection{Spectra analysis and uncertainty determination} 

The measured transmission spectrum of CO$_2$ broadened by H$_2$ can be expressed as:

\begin{multline}
T(\sigma, P_{\mathrm{CO_2}}, P_{\mathrm{H_2}}, T) 
=\\
\exp \Bigg[ 
    -L \, P_{\mathrm{CO_2}} 
    \sum_i S_i(T) \, \phi(\sigma - \sigma_i, P_{\mathrm{CO_2}}, P_{\mathrm{H_2}}, T) 
\Bigg] \;\otimes\; \mathrm{ILS}
\end{multline}
where the symbol $\otimes$ indicates the convolution between the high-resolution transmission with the instrument line-shape function, L (cm) is the absorption path length, and $P_{\mathrm{CO_2}}$ and $P_{\mathrm{H_2}}$ (atm) are respectively the CO$_2$ and H$_2$ pressures in the mixture. The sum is over all lines contributing to absorption at wavenumber $\sigma$ (cm$^{-1}$). The quantity S$_i$ (T) (cm$^{-2}$ atm$^{-1}$) is the intensity of the line $i$ at temperature T (K) and $\phi(\sigma - \sigma_i, P_{\mathrm{CO_2}}, P_{\mathrm{H_2}}, T)$ (in cm) is the line shape or the normalized profile of this line. 
In the simplest case, this line-shape function is expressed by the Voigt profile, which is a convolution of a Gaussian and a Lorentzian profile, taking into account respectively the Doppler broadening and pressure-induced broadening and shift of the line. More sophisticated models can be used for $\phi$ as well, taking into account more refined pressure-induced collisional effects such as the speed dependence of the line broadening and shift, the Dicke narrowing, and the collision-induced interferences between line (i.e., line mixing) \citep{ngo_isolated_2013, mHT_2025}. 

Here we used three different profiles to fit the measured spectra: the usual Voigt profile, the Voigt profile with line mixing (LM) accounted for, and the hard-collision (HC) profile with LM. The Doppler width was calculated for the temperature of the measurement of each spectrum and fixed in the fits. For the Voigt profile, two line-shape parameters were fitted: the H$_2$ pressure-broadening coefficient, $\gamma_{\mathrm{H_2}}$, and the H$_2$ pressure shift, $\delta_{\mathrm{H_2}}$ (both in cm$^{-1}$/atm). To model the weak LM effects between CO$_2$ lines at the considered pressures, we used the first-order approximation \citep{Rosenkranz_1975}. In this case, the LM coefficient,  $\xi$ (atm$^{-1}$), is also adjusted. Finally, to model the weak Dicke narrowing effect, due to molecule velocity changes caused by collisions, we used the hard-collision model \citep{nelkin1964simple}, leading to an additional parameter to be fitted, the Dicke narrowing parameter, or the velocity changing rate, $\nu_{\mathrm{VC}}$ (cm$^{-1}$/atm). When the signal-to-noise ratio of the measured spectra is high enough, refined line-shape models lead to better agreement with measurements. The retrieved line-broadening coefficient also slightly depends on the used line shapes (see \citealt{ngo_isolated_2013} and references therein).

A multi-fitting procedure was used, in which we fitted simultaneously the four measured spectra of CO$_2$ in H$_2$, constraining the different line-shape parameters (i.e., $\gamma_{H_2}$, $\delta_{H_2}$, $\xi$, and $\nu_{VC}$) to be the same at different pressures. The instrument line-shape function, determined from the pure CO$_2$ spectrum measured at low pressure, as explained in Sect. 2.1, was fixed in the fits. In addition to the line-shape parameters, the line area [i.e., L*$P_{CO_2}$*S$_i$ (T)] and line position (at zero pressure) were also fitted. Spectral ranges about 2 cm$^{-1}$ were considered at a time in the fits. All lines with integrated intensity greater than $1 \times 10^{-20} cm^{-1}/(molec \cdot cm^{-2})$ were fitted. These correspond to lines of the $\nu_3$ band of the main isotopologue of CO$_2$ with rotational quantum number $J$ up to 62. The contribution of weaker lines was calculated for the measured experimental conditions using spectroscopic data from the HITRAN database \citep{gordon_hitran2020_2021}, except for the line broadening coefficient. The latter was estimated from the first round fits of the $\nu$$_3$ lines of $^{12}$CO$_2$ and neglecting any vibrational and isotopic dependences of the broadening coefficient. For the calculation of weak lines, the partial pressures of CO$_2$ for each spectrum was redetermined using the fitted areas and integrated intensities in HITRAN for several strong $\nu$$_3$ lines of $^{12}$CO$_2$. For each spectral range and pressure, we used two baseline models for the 100\% transmission: a linear baseline and a third-order polynomial function. The parameters of the baseline were also fitted. The difference in the retrieved line-shape parameters obtained with these two baseline models was accounted for in the estimation of their uncertainties. 

\begin{figure}[htbp]
  \centering
  \includegraphics[width=0.5\textwidth]{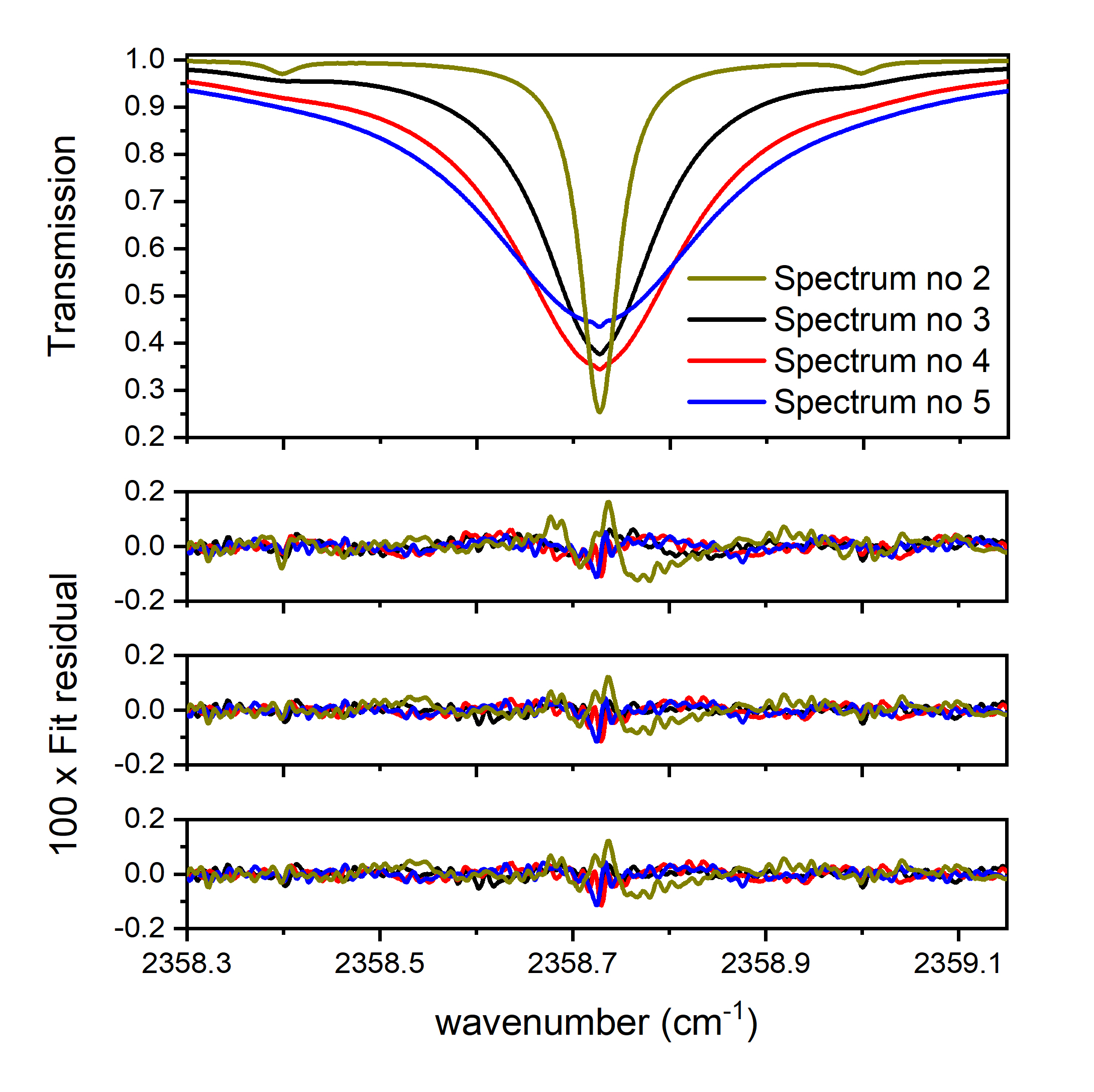}
  \caption{Measured transmission of the R(12) line (top panel) and fit residuals obtained by fitting the measured spectra with the Voigt profile (second panel), with LM accounted for (third panel) and with the HC model together with LM (bottom panel).}
  \label{fig:fit_exp}
\end{figure}

Examples of the measured transmissions and the corresponding fit residuals obtained with the three line-shape models are shown in Fig.~\ref{fig:fit_exp}. The residuals obtained with LM taken into account are close to the experimental noise level; the weak remaining residuals are probably due to the imperfect modeling of the instrument line shape. It was not possible to fit all lines with the HC model with first-order LM, probably due to the limited signal-to-noise ratio of the measured spectra. This model resulted in only marginally better fit residuals compared to those obtained with the Voigt model with LM. In addition, the differences between the $\gamma_{\mathrm{H_2}}$ values obtained with Voigt+LM and HC+LM are well below the total uncertainty. We note that a similar conclusion was obtained when using the speed-dependent Voigt profile \citep{ngo_isolated_2013} to fit the measured spectra. Therefore, in the following, we focus and discuss results obtained with Voigt+LM only.   

The uncertainty of the fitted parameters comes from different sources. The first is the statistical uncertainty (type A), corresponding to the standard deviation obtained from the fits. This quantity is negligibly small compared to the systematic uncertainties. The latter arise from the uncertainty in the measured pressure and temperature, and in the apparatus function. In our analysis, we assumed that self-broadening effects are the same as H$_2$ broadening effects, i.e., $P_{H_2}=P_{tot}$. This assumption has a largely negligible effect on the determined H$_2$ collision-induced line-shape parameters since the concentrations of CO$_2$ in the mixtures are low (see Table~\ref{Table1}). To determine the uncertainty due to errors in the measured $P_{tot}$, we reanalyzed the four measured spectra, but with the total pressure $P_{\text{tot}} = P_{\text{tot}}^{\text{meas}} + \Delta P$, where $\Delta$P is the error in the measured total pressure. The corresponding uncertainty of the line-shape parameters was then determined as the difference between the parameters obtained using ($P_{\text{tot}}^{\text{meas}}$+$\Delta$ P) and those obtained using $P_{\text{tot}}^{\text{meas}}$. A similar procedure was performed to determine the uncertainty due to the uncertainty of the measured temperature. In addition, the line-shape parameters were considered as being obtained at the average temperature. The uncertainty due to the difference between the average temperature and the temperature measured for each spectrum was also considered, using the temperature dependence exponent of H$_2$ broadening coefficients provided in the HITRAN database \citep{gordon_hitran2020_2021}. Uncertainties due to errors in the ILS were estimated by performing fits with two ILSs, both determined using LINEFIT, by applying LINEFIT to selected microwindows and to the entire R branch in the pure CO$_2$ spectrum. The total uncertainty for each retrieved parameter was then computed as the sum of all individual contributions, including that arising from the choice of baseline models. Except for a few lines of high rotational quantum number, this computed uncertainty is below 2\%. We therefore assume that the total uncertainty of the measured $\gamma_{\mathrm{H_2}}$ for these lines is 2\%.    

\subsection{Results and comparison with available data}
\begin{figure}[htbp]
  \centering
  \includegraphics[width=0.5\textwidth]{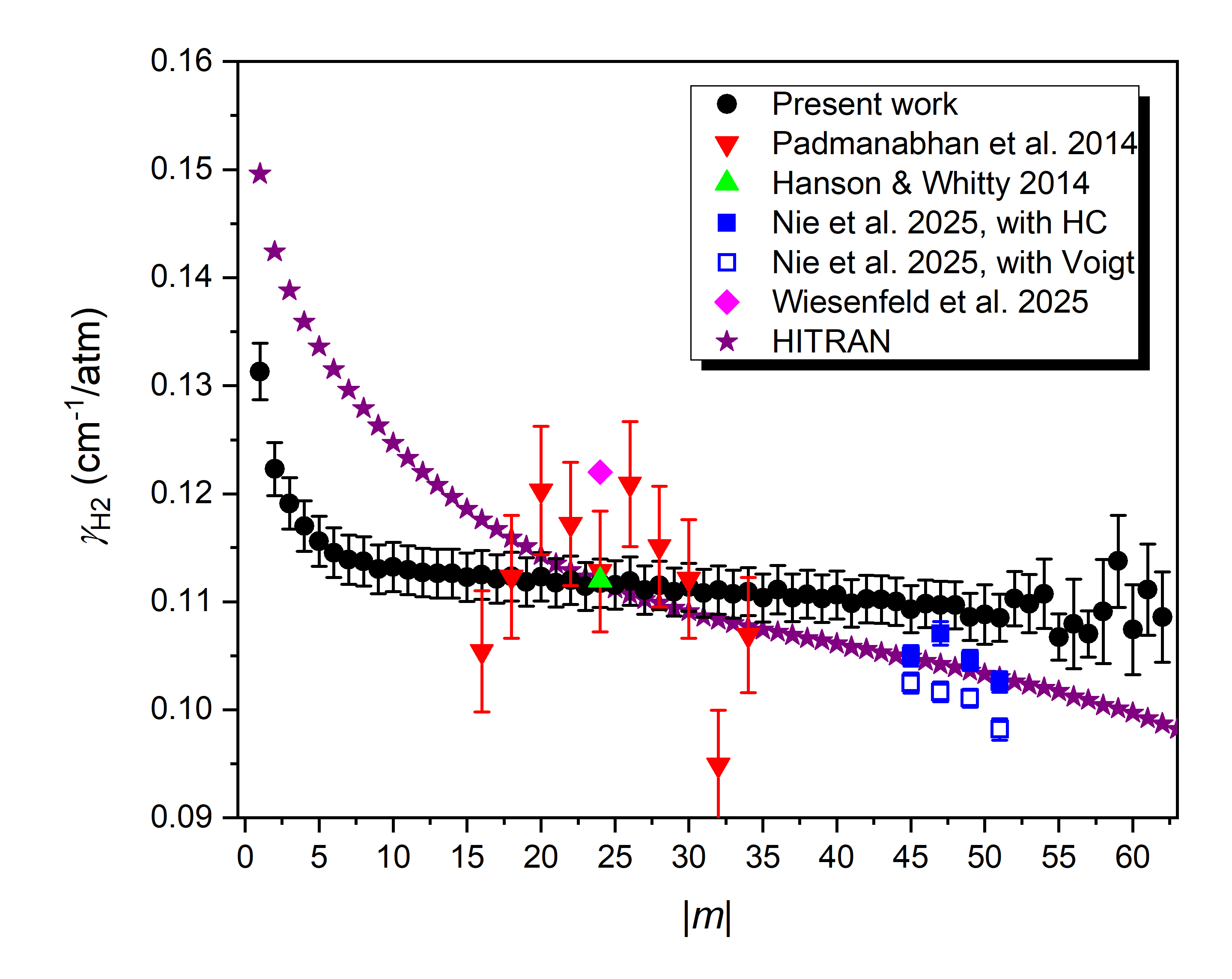}
  \caption{H$_2$ broadening coefficient measured in this work, and a comparison with the available experimental and theoretical data and with the values from HITRAN \citep{tan_h2_2022}.}
  \label{fig:broadening}
\end{figure}

The obtained values of $\gamma_{\mathrm{H_2}}$ are listed in \tab{Table2}, and are plotted in Fig.~\ref{fig:broadening} as a function of $\left| m \right|$, where $m$ = -$J$ and $m$ = $J$+1 in the P and R branches, respectively.  
\begin{table*}[t]
	\centering
    \caption{H$_2$ pressure-broadening and pressure-shift coefficients of CO$_2$ lines in the $\nu_3$ band measured around 294 K. }
	\begin{tabular}{r c c c r c c }
        \hline  \hline 
        $m$ & $\gamma_{\mathrm{H_2}}$ &  $\delta_{\mathrm{H_2}}$ &   & $m$ & $\gamma_{\mathrm{H_2}}$ &  $\delta_{\mathrm{H_2}}$\\
        \hline 
        -62 & 0.1086 (42)  &    -      &   & 1 & 0.1313 (26)  & -0.0045 (6)         \\
        -60 & 0.1074 (41)  &   -       &   & 3 & 0.1191 (24)  & -0.0035 (6)         \\   
        -58 & 0.1091 (48)  &  -        &   & 5 & 0.1156 (23)  & -0.0047 (4)         \\  
        -56 & 0.1079 (42)  &  -        &   &   7 & 0.1139 (23)  & -0.0036 (4)         \\   
        -54 & 0.1107 (32)  & -         &   & 9 & 0.1130 (23)  & -0.0032 (6)         \\   
        -52 & 0.1103 (25)  & -0.0047 (10) &  &    11 & 0.1129 (23)  & -0.0031 (7)         \\    
        -50 & 0.1088 (27)  & -         &  &   13 & 0.1126 (23)  & -0.0025 (3)         \\   
        -48 & 0.1097 (25)  & -0.0050 (11) & &  15 & 0.1123 (22)  & -0.0029 (6)         \\ 
        -46 & 0.1098 (22)  & -0.0049 (5)  &  &   17 & 0.1121 (22)  & -0.0031 (6)         \\  
        -44 & 0.1100 (22)  & -0.0045 (6)   &  &  19 & 0.1125 (22)  & -0.0032 (8)         \\     
        -42 & 0.1102 (22)  & -0.0048 (6)    &  & 21 & 0.1117 (22)  & -0.0031 (6)         \\    
        -40 & 0.1106 (22)  & -0.0042 (6)    &  & 23 & 0.1114 (22)  & -0.0025 (4)         \\     
        -38 & 0.1107 (22)  & -0.0042 (4)   &  &  25 & 0.1115 (22)  & -0.0031 (6)         \\
        -36 & 0.1111 (22)  & -0.0046 (4)   &  & 27 & 0.1110 (22)  & -0.0032 (7)         \\
        -34 & 0.1109 (22)  & -0.0041 (6)   &  & 29 & 0.1109 (22)  & -0.0032 (7)         \\   
        -32 & 0.1111 (22)  & -0.0044 (6)   &  &  31 & 0.1108 (22)  & -0.0028 (4)         \\ 
        -30 & 0.1113 (22)  & -0.0043 (4)   &  &   33 & 0.1107 (22)  & -0.0031 (6)         \\    
        -28 & 0.1115 (22)  & -0.0047 (4)   &  &  35 & 0.1103 (22)  & -0.0034 (7)         \\  
        -26 & 0.1119 (22)  & -0.0053 (5)   &  & 37 & 0.1104 (22)  & -0.0033 (6)         \\  
        -24 & 0.1117 (22)  & -0.0043 (4)    &  &  39 & 0.1103 (22)  & -0.0029 (3)         \\ 
        -22 & 0.1120 (22)  & -0.0041 (6)    &  & 41 & 0.1099 (22)  & -0.0033 (7)         \\  
        -20 & 0.1123 (22)  & -0.0044 (4)   &  &  43 & 0.1102 (22)  & -0.0035 (6)         \\
        -18 & 0.1123 (22)  & -0.0049 (6)   &  &  45 & 0.1093 (22)  & -0.0029 (4)         \\  
        -16 & 0.1125 (22)  & -0.0042 (6)   &  &  47 & 0.1097 (22)  & -0.0034 (5)         \\   
        -14 & 0.1126 (23)  & -0.0045 (7)   &  & 49 & 0.1086 (22)  & -0.0034 (8)         \\  
        -12 & 0.1127 (23)  & -0.0042 (6)   &  &  51 & 0.1085 (22)  & -0.0042 (7)         \\   
        -10 & 0.1132 (23)  & -0.0047 (6)   &  &  53 & 0.1098 (27)  & -0.0031 (7)         \\ 
        -8 & 0.1137 (23)  & -0.0045 (8)   &  & 55 & 0.1067 (21)  & -0.0039 (11)         \\ 
        -6 & 0.1145 (23)  & -0.0042 (8)  &  &   57 & 0.1070 (21)  & -0.0044 (3)         \\ 
        -4 & 0.1170 (23)  & -0.0041 (6)    &  &  59 & 0.1137 (43)  & -0.0028 (24)         \\  
        -2 & 0.1223 (24)  & -0.0058 (9)    &  &     61 & 0.1111 (42)  & -         \\   
   
        \hline 
	\end{tabular}

 \label{Table2}
 
\tablefoot{
\tablefoottext{a}{These coefficients are expressed in cm$^{-1}$/atm. The number in parentheses indicates the corresponding uncertainty and has the same unit as the last significant digit.}
}
\end{table*} 
The data available in the literature are also displayed in Fig.~\ref{fig:broadening} for comparison. As can be observed, for the P(24) line, our result is in very good agreement with that of \citet{Hanson_Whitty_2014}, with a difference of 0.25\%. The calculated value of \citet{Wiesenfeld_2025} for the same line is about 9\% larger than our result. Larger discrepancies, up to 15\%, are observed when comparing our values with those of \citet{PADMANABHAN_JQSRT_2014}. In the latter study, a large and apparently random variation of $\gamma_{\mathrm{H_2}}$ with $\left| m \right|$ is observed, which is not the case for our data. Differences of up to 9\% and 5\% are observed when comparing our results with those of \citet{NIE_2025} obtained with the Voigt and the HC profiles, respectively. We note that \citet{NIE_2025} determined line intensity and broadening coefficients from spectra measured for mixtures of CO$_2$ with various perturbers such as O$_2$, N$_2$, H$_2$, and Ar. Their retrieved line intensity depends strongly on the perturber, with variation up to 3.2\%, indicating large uncertainty in the retrieved parameters. Therefore, the actual uncertainty in the measured broadening must be significantly larger than the 1\% uncertainty claimed. Finally, significant differences of up to 16.5\% are also found when comparing our results with the values provided in HITRAN \citep{tan_h2_2022}. The $\gamma_{\mathrm{H_2}}$ values in HITRAN were obtained by simply scaling the air-broadening coefficient to match the results of \citet{Hanson_Whitty_2014} at $\left| m \right|$ = 24 \citep{tan_h2_2022}. Our results therefore demonstrate that the rotational dependence of $\gamma_{\mathrm{H_2}}$ differs significantly from that of $\gamma_{\mathrm{air}}$. Consequently, using these data \citep{PADMANABHAN_JQSRT_2014, tan_h2_2022} for opacity modeling of planetary and exoplanetary atmospheres may lead to substantial errors. 

\begin{figure}[htbp]
  \centering
  \includegraphics[width=0.5\textwidth]{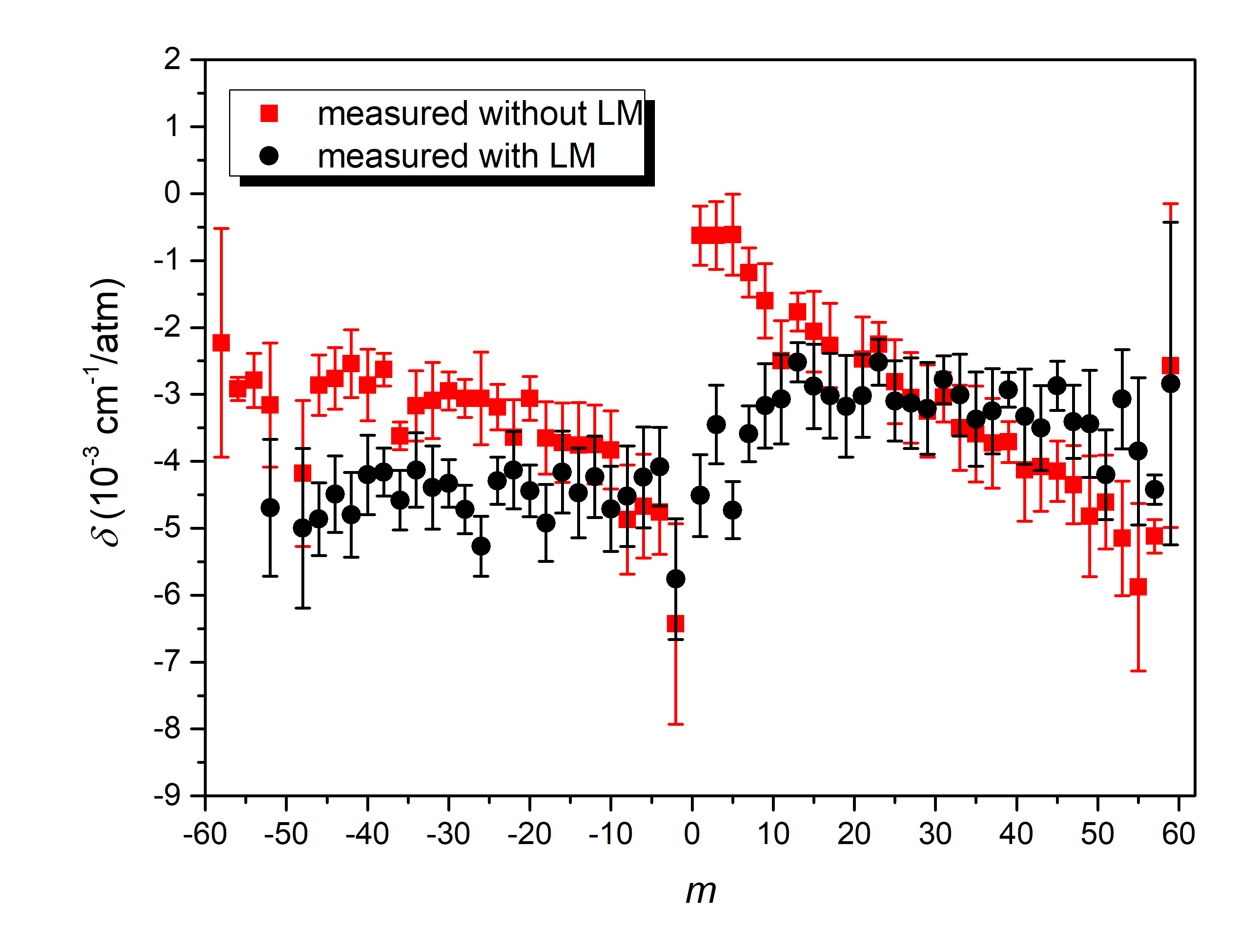}
  \caption{H$_2$ pressure shift determined from fits of measured spectra with the Voigt profile (without LM) and with LM taken into account.}
  \label{fig:shift}
\end{figure}

The measured pressure-shift coefficients, $\delta_{H_2}$, are plotted in Fig.~\ref{fig:shift} versus the rotational quantum number. This work provides the first determination of $\delta_{\mathrm{H_2}}$ for the $\nu_3$ band. It is important to emphasize that $\delta_{\mathrm{H_2}}$ were obtained by fitting the measured spectra accounting for LM effects. When a simple Voigt profile is used (neglecting LM), the derived coefficients differ significantly and exhibit a strong rotational dependence (see Fig. ~\ref{fig:shift}). This behavior is due to the influence of line-mixing effects, which can shift the peak absorption. As can be seen, the rotational dependence of $\delta_{\mathrm{H_2}}$ is weak, similarly to what was observed for CO$_2$ broadened by He \citep{HENDAOUI_CMDS_CO2He_2026}, although the magnitude of $\delta_{\mathrm{H_2}}$ is significantly larger in the case of H$_2$. The mean value of $\delta_{\mathrm{H_2}}$ is -3.9 10$^{-3}$ cm$^{-1}$/atm, compared to -0.52 10$^{-3}$ cm$^{-1}$/atm for He. To the best of our knowledge, the only previously available data on $\delta_{\mathrm{H_2}}$ are those reported by \citet{NIE_2025}. In that study, H$_2$ pressure-induced shifts were measured for four CO$_2$ lines in the $\nu_1+2\nu_2+\nu_3$ band, yielding a mean value of about -7.5 10$^{-3}$ cm$^{-1}$/atm. The difference between their result and the present measurements can be attributed to the strong vibrational dependence of the pressure-induced line shift.

\section{Theoretical predictions of $\gamma_{H_2}$}\label{sec:stateoftheart}

\subsection{Calculation method and data used}
We employed rCMDSs to predict $\gamma_{H_2}$ over a broad temperature range, from 200 K to 1000 K. Using accurate intermolecular potentials and the Newtonian equations of motion, this method enables the simulation of the time evolution of a system containing a large number of molecule at a given pressure and temperature \citep{Allen-Tidesley}. The auto-correlation functions (ACFs) of the dipole moment, responsible for the transitions, can thus be computed. The Fourier-Laplace transform of these ACFs directly yields the spectral density or the normalized spectral profile. In addition, we applied a requantization procedure \citep{Hartmann2013a} to the molecular rotation. The requantized line spectrum obtained this way can then be fitted with a line-shape model, such as the usual Voigt profile, to deduce the corresponding pressure broadening. This approach has enabled very satisfactory predictions of line-shape parameters \citep{Nguyen_CO2N2_JCP_2018, TRAN_CMDS_O2_2019, NGUYEN_CMDS_CO2_2020, NGO_CMDS_N2O_2021, HENDAOUI_CMDS_CO_2024, NGO_Tran_CMDS_CO2_2025, TRAN_CMDS_CO2N2_2025, HENDAOUI_CMDS_CO2He_2026}, far-wing absorption \citep{Hartmann_JCP_2010, Hartmann_CMDS_CO2H2O_2018}, and collision-induced absorption \citep{Hartmann_CIA_CO2_2011, Hartmann_CMDS_CIA_N2_2017, FAKHARDJI_CIA_CH2CO2_2022} for various molecular systems and under wide ranges of pressure and temperature. 

At the initial time, the molecules were positioned as follows: the center of mass of each molecule was randomly assigned, with a minimum separation of 7 \AA\ to prevent unphysically strong interactions. The system then reaches equilibrium after a temperature- and pressure-dependent temporization time, \( t_{\text{tempo}} \). Translational and angular speeds were initialized according to Maxwell--Boltzmann distributions, with random velocity vectors and molecular axis orientation assigned to each molecule. At each time step, the forces and torques acting on each molecule were computed to update their accelerations, translational velocities, position, angular velocities, and molecular orientations at the next time step \( t + \Delta t \).

A requantization procedure \citep{Hartmann2013a} was applied to the classical rotation of the CO$_2$ molecules, but only when the interaction between the considered CO$_2$ molecule and its neighboring H$_2$ molecules was negligible. This approach, well suited for the CO$_2$ molecule due to the small energy separation between its rotational levels, was shown to yield excellent predictions of CO$_2$ broadening coefficients in N$_2$, O$_2$, He and self-broadening \citep{Nguyen_CO2N2_JCP_2018, NGUYEN_CMDS_CO2_2020, NGO_Tran_CMDS_CO2_2025, HENDAOUI_CMDS_CO2He_2026}. Specifically, for a molecule of rotational angular speed \( \omega \), the even integer \( J \) is determined such that \( \frac{\hbar J}{I} \) is closest to \( \omega \), where \( I \) is the moment of inertia. This procedure matches the classical rotational frequency with the quantum position of the P-branch lines. Once \( J \) is found, \( \omega \) is requantized by applying the change \( \omega = \frac{\hbar J}{I} \), while the orientation of the rotational angular momentum remains unchanged. This P-branch requantization implies that the R-branch will be an exact mirror image of the P-branch. 

The ACF of the dipole moment vector, assumed along the CO$_2$ molecular axis (as is the case for the asymmetric stretching absorption bands of CO$_2$), was then calculated during the rCMDSs. The Doppler effect associated with the translational motion was also included. Finally, the spectral profile (or the normalized absorption coefficient) was obtained from the Fourier--Laplace transform of the dipole moment ACF \citep{hartmann_collisional_2021}. 

Here CO$_2$ molecules were modeled as linear rigid rotors in their equilibrium configuration, with all effects of vibrational motion disregarded in the calculations. This approximation is expected to have a negligible effect on the calculated pressure broadening \citep{HASHEMI_CO2_JQSRT_2020}. The CO$_2$-H$_2$ interaction was described using the site-site potential provided by \citet{Hellmann_CO2H2_PES_2025}. The parameters of this site-site potential, as well as the positions of the sites in the molecules were determined from the ab initio  calculated interaction energies as described in \citet{Hellmann_CO2H2_PES_2025}. The spectra were thus calculated without using any parameter adjusted to the experimental data. We note that our rCMDSs do not account for dipole dephasing, which arises from different intermolecular interactions in the upper and lower states of the transitions. As a result, the calculated spectra do not show any pressure-induced vibrational shifts.

The simulations were performed on the Jean-Zay HPE SGI 8600 supercomputer at the Institut du Dévéloppement et des Ressources en Informatique Scientifique, which supports massively parallel computing. The molecules were distributed across four thousand boxes with periodic boundary conditions \citep{Allen-Tidesley}, each containing 20 000 molecules. Mixtures containing 50\% CO$_2$ were considered, but only CO$_2$-H$_2$ interactions were accounted for. This is equivalent to CO$_2$ being infinitely diluted in H$_2$, with the total pressure corresponding to the partial pressure of H$_2$. rCMDSs were performed for the following temperatures: 200 K, 296 K, 400 K, 500 K, 600 K, 800 K, and 1000 K. For each temperature, the pressure of the mixture was chosen such that  line-mixing effects were negligible (pressures up to 0.5 atm for $T$ $\leq$ 400 K and up to 1 atm for $T$ between 500 K and 1000 K). For each temperature, a total number of \(8 \times 10^7\) molecules was considered, yielding signal-to-noise ratios of the calculated spectra of up to 1000 for the most intense lines. 

Spectra calculated with rCMDS were fitted with Voigt profiles to retrieve the pressure-broadening coefficients. The line position, line area, and a linear baseline were also fitted, as for the measured spectra. Additional fits were performed using more refined models, such as the speed-dependent Voigt and HC profiles, to account for the speed dependence of line broadening and for Dicke narrowing, respectively. However, the results show that these effects have a negligible influence on the fit residuals and on the retrieved line-broadening coefficients. The pressure-broadening coefficients obtained from the rCMDS-calculated spectra at all considered temperatures are presented and compared with the experimental values in the next subsection.

\subsection{H$_2$ broadening coefficient predicted by rCMDS}
\begin{figure}[htbp]
  \centering
  \includegraphics[width=0.5\textwidth]{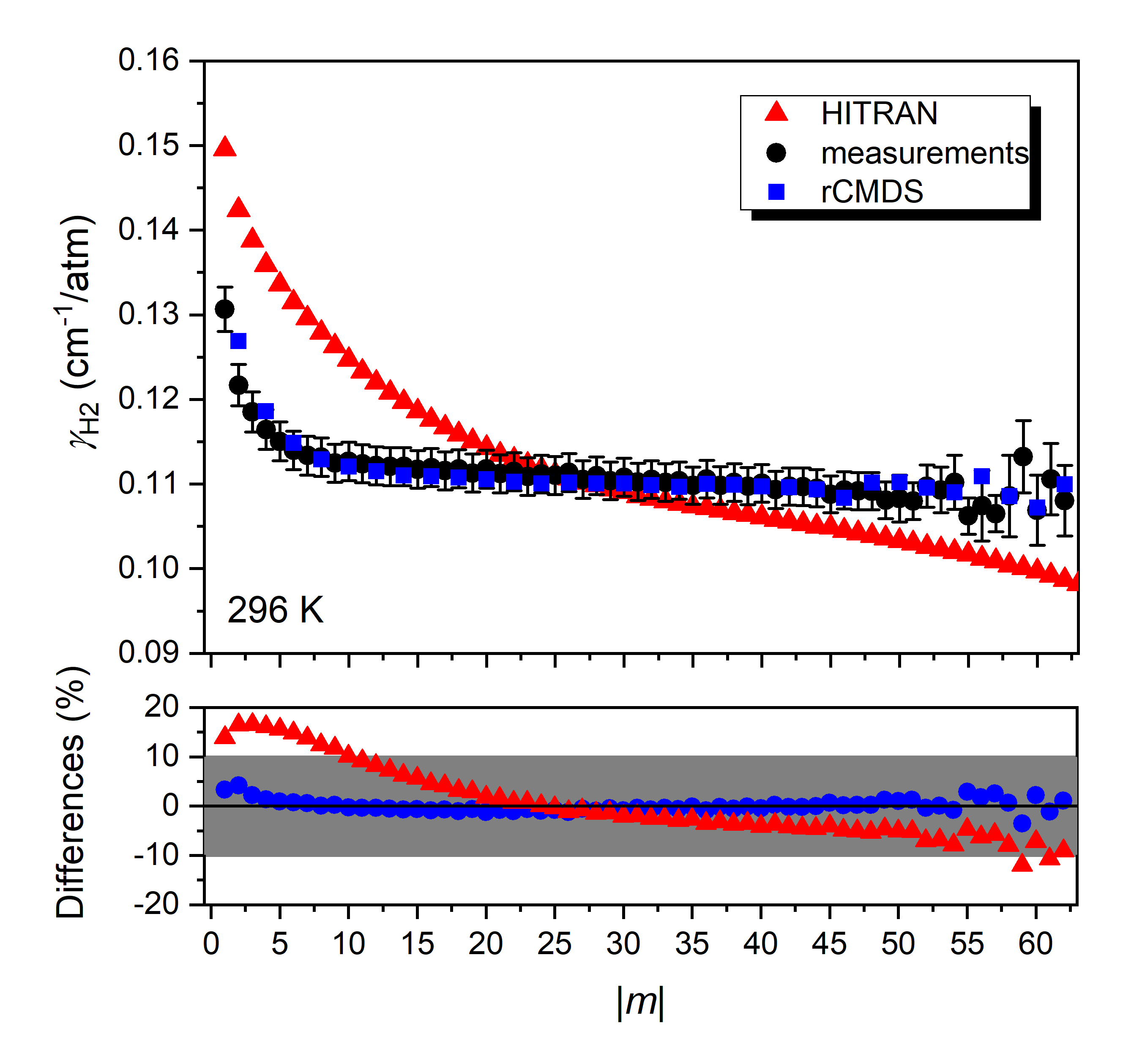}
  \caption{H$_2$ broadening coefficient predicted by rCMDS at 296 K, and a comparison with the experimental values of the present work and with the HITRAN data. The gray space represents the 10\% precision level required for exoplanetary atmosphere studies.}
  \label{fig:broadening_CMDS_296K}
\end{figure}

Figure~\ref{fig:broadening_CMDS_296K} shows the broadening coefficients obtained from rCMDS at 296 K and their comparison with our accurate experimental values, where a very good agreement can be observed. Except for the P(2) line, for which a difference of 4\% is observed, the discrepancies between the rCMDS predictions and the measured values are well below the experimental uncertainty (2\%). We note that, for this comparison, the measured values were converted to 296 K using a single power law and the associated temperature dependence obtained from rCMDS, as listed in Appendix~\ref{appendix}. This level of agreement is significantly better than the precision of 10\% required for opacity calculations in exoplanet studies with JWST \citep{Wiesenfeld_2025}. In contrast, the current HITRAN data (red curve in Fig.~\ref{fig:broadening_CMDS_296K}, \citep{tan_h2_2022}) leads to substantial deviations of up to 16.5\% relative to the rCMDS values. The comparison presented in Fig. ~\ref{fig:broadening_CMDS_296K} therefore fully confirms the validity of our rCMDS method for predicting H$_2$ broadening coefficient of CO$_2$ lines.

\begin{figure*}[h]
  \centering
  \includegraphics[width=\textwidth]{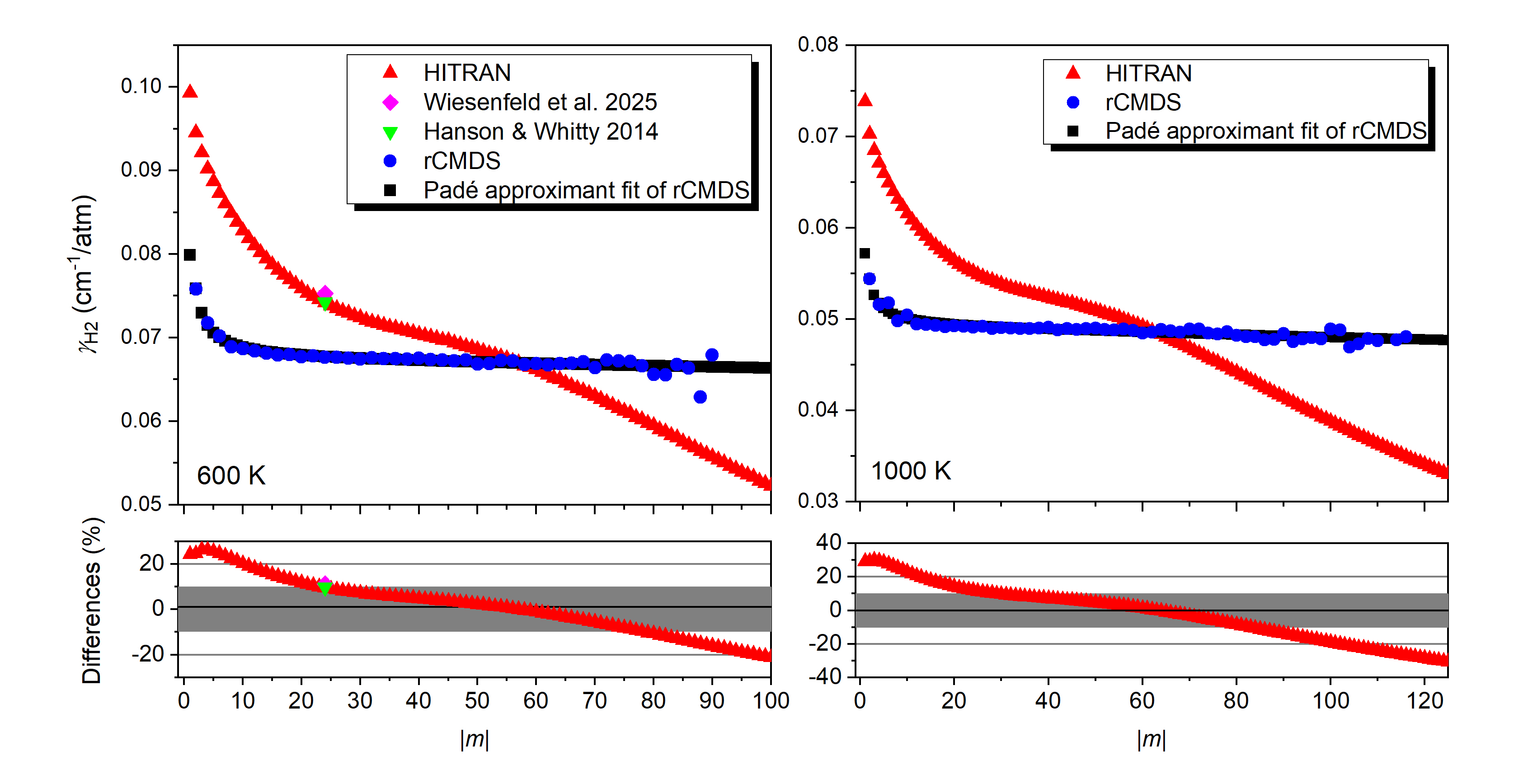}
  \caption{H$_2$ broadening coefficient predicted by rCMDS at 600 K and 1000 K, and a comparison with values of HITRAN \citep{tan_h2_2022}. The gray space represents the 10\% precision level required for exoplanetary atmosphere studies.}
  \label{fig:CMDS_T}
\end{figure*}

rCMDS calculations were performed for six additional temperatures: 200 K, 400 K, 500 K, 600 K, 800 K, and 1000 K. For each temperature, $\gamma_{\mathrm{H_2}}$ was carefully determined for all considered lines. As the number of populated rotational levels increases with temperature, $\gamma_{\mathrm{H_2}}$ was determined up to a maximum rotational quantum number $J$ = 50 at 200 K, while at 1000 K line broadening was obtained for lines up to $J$ = 120. Figure ~\ref{fig:CMDS_T} presents the rCMDS-predicted results at 600 K and 1000 K, together with corresponding $\gamma_{\mathrm{H_2}}$ values from HITRAN \citep{tan_h2_2022}. Large discrepancies are observed at all temperatures, with deviations reaching more than 30\%. 

At 600 K, our results are also compared with the experimental measurements of \citet{Hanson_Whitty_2014}, who measured the H$_2$ broadening coefficient for the P(24) line and provided a temperature dependence exponent of 0.582. The value calculated by \citet{Wiesenfeld_2025} at 600 K for the same line is also shown. As illustrated in Fig.~\ref{fig:CMDS_T}, our rCMDS results are about 10\% smaller than those of \citet{Hanson_Whitty_2014} and \citet{Wiesenfeld_2025}. Nevertheless, given the excellent agreement between rCMDS predictions and accurate experimental data at room temperature (Fig.~\ref{fig:broadening_CMDS_296K}), as well as the improved validity of classical treatment for molecular rotations at higher temperatures, we are confident in the reliability of our results. Additional accurate measurements of H$_2$ broadening coefficients at temperatures above room temperature would nevertheless be valuable for further validation.  

As can be seen in Figs.~\ref{fig:broadening_CMDS_296K},~\ref{fig:CMDS_T}, a noticeable scatter is present in the rCMDS-predicted broadening coefficients, particularly for high |$m$| values. 
This scatter mainly arises from the lower signal-to-noise ratio of the corresponding lines, which leads to increased uncertainty in the extracted line broadening. To mitigate this effect, the rCMDS-predicted broadening coefficients at each temperature were fitted using a Padé approximant, as was done in \citet{tan_h2_2022, HENDAOUI_CMDS_CO2He_2026}. As illustrated by the examples in Fig. ~\ref{fig:CMDS_T}, the quality of these fits is excellent. The fitted functions were then used to deduce broadening coefficients for R-branch lines, assuming symmetry between P- and R-branch coefficients with respect to $m$. This symmetry assumption has a negligible impact on the collisional line broadening of CO$_2$ \citep{MONDELAIN2025109271}. In addition, as $\gamma_{\mathrm{H_2}}$ tends toward a constant value at high $J$ (see Figs.~\ref{fig:broadening_CMDS_296K},~\ref{fig:CMDS_T}), these functions can be safely used to extrapolate $\gamma_{\mathrm{H_2}}$ for $J$ up to 120 for all considered temperatures. The fitted broadening coefficients were then used to determine their temperature dependence, which is presented in the following subsection.  

\subsection{Temperature dependence}
We used both the usual single power law (SPL) and the recently recommended double power law \citep{gamache2018temperature} (DPL) to model the temperature dependence of $\gamma_{\mathrm{H_2}}$. In the SPL, the temperature dependence of $\gamma_{\mathrm{H_2}}$ is expressed as 

\begin{equation}
\gamma_{\mathrm{H_2}}(T)=\gamma_{\mathrm{H_2}}(T_0)\left(\frac{T_0}{T}\right)^n,
\end{equation}

with T$_0$ = 296 K and $n$ the temperature-dependence exponent. In the DPL, $\gamma_{\mathrm{H_2}}(T)$ is written as
\begin{equation}
    \gamma_{\mathrm{H_2}}(T)=\gamma_{0}\left(\frac{T_{0}}{T}\right)^{n_{1}}+\gamma_{0}^{\prime}\left(\frac{T_{0}}{T}\right)^{n_{2}}.
\end{equation}

For each transition, both the SPL and DPL were fitted to the rCMDS-predicted values of $\gamma_{\mathrm{H_2}}$ at six temperatures: 200 K, 296 K, 400 K, 600 K, 800 K, and 1000 K. In the SPL fits the parameters $\gamma_{\mathrm{H_2}}(T_0)$ and $n$ were fitted, while in the DPL four parameters $\gamma_{0}$, $\gamma_{0}^{\prime}$, $n_1$ and $n_2$ were fitted.

\begin{figure}[htbp]
  \centering
  \includegraphics[width=0.5\textwidth]{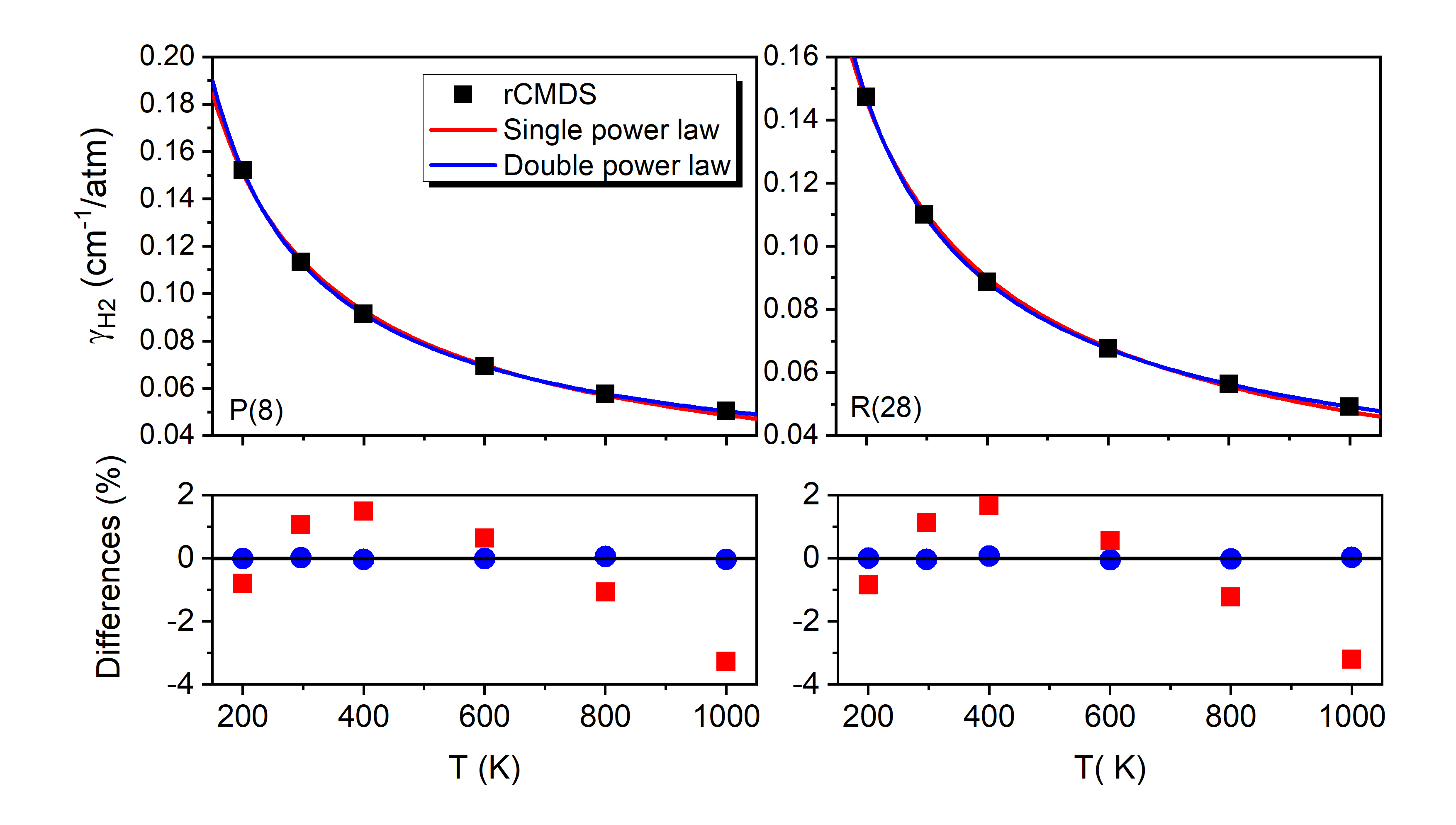}
  \caption{rCMDS-predicted H$_2$ broadening coefficients vs. temperature (black squares) for two lines P(8) and R(28), and their fits using a single power function (red line) and a double power function (blue line). The differences between the rCMDS values and fits are displayed in the corresponding lower panels.}
  \label{fig:temperature dependence}
\end{figure}

Figure~\ref{fig:temperature dependence} presents examples of fits with the SPL and DPL for the P(8) and R(28) lines. The corresponding differences between the rCMDS results and the fitted broadenings are also shown. As can be observed, the DPL provides better overall agreement with rCMDS predictions, with deviations always below 1\%, whereas the SPL leads to discrepancies of up to 4\%, especially at high temperatures. 

Examples of the fitted parameters ($\gamma_{\mathrm{H_2}}(T_0)$, $n$, $\gamma_{0}$, $\gamma_{0}^{\prime}$, $n_1$, and $n_2$), together with their standard deviations are reported in Appendix~\ref{appendix}, while the results for all considered lines are available on the Zenodo platform at the following link: 10.5281/zenodo.18350127. To further verify the quality of the fits, we recalculated $\gamma_{H_2}$ using both the SPL and DPL parameterizations and compared the results with direct rCMDS predictions at 500 K. The results show very good agreement, with  maximum deviations of 3\% for the SPL and 1\% for the DPL. 

\section{Conclusions}\label{sec:stateoftheart}
Accurate and comprehensive H$_2$ broadening coefficients for CO$_2$ lines under a wide temperature range are necessary for reliable opacity modeling of exoplanetary atmospheres. In this work, we reported the first accurate measurements at room temperature of H$_2$ broadening and pressure-shift coefficients for the entire CO$_2$ band at 4.3 $\mu$m, obtained using a high-resolution Fourier transform spectrometer. The commonly used data derived from air-broadening coefficients scaled by a constant factor show significant discrepancies, up to 16.5\%, compared to the measured coefficients.\\
In addition, we performed first-principle calculations based on requantized classical molecular dynamics simulations (rCMDSs) to predict H$_2$ broadening coefficients of CO$_2$ lines over a wide temperature range, from 200 K to 1000 K. The predicted values are in very good agreement with the experimental measurements; the maximum differences are well below 3\%, thus satisfying the precision requirements for JWST exoplanet studies. The experimentally validated rCMDS results were then used to derive the temperature dependence of the line broadening.\\
The obtained dataset allows the calculations of H$_2$ broadening coefficients for CO$_2$ lines with rotational quantum number up to 120 at any temperature within the 200--1000 K range. These results are expected to significantly improve the accuracy of opacity calculation for H$_2$-rich atmosphere exoplanets. 

\section*{Data availability}
The complete dataset produced in this work are accessible at the following Zenodo link: 10.5281/zenodo.18350127

\begin{acknowledgements}
This work was granted access to the HPC resources of IDRIS under the allocation 2025-[A0180914906] made by GENCI. It was also supported by the Programme National de Planétologie (PNP) of CNRS-INSU co-funded by CNES.
\end{acknowledgements}

\bibliographystyle{aa}
\bibliography{biblio}

@article{gordon_hitran2020_2021,
	title = {The {HITRAN2020} molecular spectroscopic database},
	issn = {0022-4073},
	url = {https://www.sciencedirect.com/science/article/pii/S0022407321004416},
	doi = {10.1016/j.jqsrt.2021.107949},
	language = {en},
	urldate = {2021-09-30},
	journal = {J. Quant. Spectrosc. Radiat. Transf.},
	author = {Gordon, I. E. and Rothman, L. S. and Hargreaves, R. J. and Hashemi, R. and Karlovets, E. V. and Skinner, F. M. and Conway, E. K. and Hill, C. and Kochanov, R. V. and Tan, Y. and Wcis{\l }o, P. and Finenko, A. A. and Nelson, K. and Bernath, P. F. and Birk, M. and Boudon, V. and Campargue, A. and Chance, K. V. and Coustenis, A. and Drouin, B. J. and Flaud, J. {\textendash}M. and Gamache, R. R. and Hodges, J. T. and Jacquemart, D. and Mlawer, E. J. and Nikitin, A. V. and Perevalov, V. I. and Rotger, M. and Tennyson, J. and Toon, G. C. and Tran, H. and Tyuterev, V. G. and Adkins, E. M. and Baker, A. and Barbe, A. and Can{\`e}, E. and Cs{\'a}sz{\'a}r, A. G. and Dudaryonok, A. and Egorov, O. and Fleisher, A. J. and Fleurbaey, H. and Foltynowicz, A. and Furtenbacher, T. and Harrison, J. J. and Hartmann, J. {\textendash}M. and Horneman, V. {\textendash}M. and Huang, X. and Karman, T. and Karns, J. and Kassi, S. and Kleiner, I. and Kofman, V. and Kwabia{\textendash}Tchana, F. and Lavrentieva, N. N. and Lee, T. J. and Long, D. A. and Lukashevskaya, A. A. and Lyulin, O. M. and Makhnev, V. Yu. and Matt, W. and Massie, S. T. and Melosso, M. and Mikhailenko, S. N. and Mondelain, D. and M{\"u}ller, H. S. P. and Naumenko, O. V. and Perrin, A. and Polyansky, O. L. and Raddaoui, E. and Raston, P. L. and Reed, Z. D. and Rey, M. and Richard, C. and T{\'o}bi{\'a}s, R. and Sadiek, I. and Schwenke, D. W. and Starikova, E. and Sung, K. and Tamassia, F. and Tashkun, S. A. and Auwera, J. Vander and Vasilenko, I. A. and Vigasin, A. A. and Villanueva, G. L. and Vispoel, B. and Wagner, G. and Yachmenev, A. and Yurchenko, S. N.},
	month = sep,
	year = {2021},
	pages = {107949},
}

@article{tan_h2_2022,
	title = {H2, {He}, and {CO2} {Pressure}-induced {Parameters} for the {HITRAN} {Database}. {II}. {Line} {Lists} of {CO2}, {N2O}, {CO}, {SO2}, {OH}, {OCS}, {H2CO}, {HCN}, {PH3}, {H2S}, and {GeH4}},
	volume = {262},
	issn = {0067-0049},
	url = {https://dx.doi.org/10.3847/1538-4365/ac83a6},
	doi = {10.3847/1538-4365/ac83a6},
	language = {en},
	number = {2},
	urldate = {2022-12-21},
	journal = {ApJ Supplement Series},
	author = {Tan, Yan and Skinner, Frances M. and Samuels, Shanelle and Hargreaves, Robert J. and Hashemi, Robab and Gordon, Iouli E.},
	month = sep,
	year = {2022},
	note = {Publisher: The American Astronomical Society},
	pages = {40},
}

@book{hartmann_collisional_2021,
	title = {Collisional {Effects} on {Molecular} {Spectra}: {Laboratory} {Experiments} and {Models}, {Consequences} for {Applications}.},
	shorttitle = {Collisional {Effects} on {Molecular} {Spectra}},
	url = {https://ui.adsabs.harvard.edu/abs/2021cems.book.....H},
	urldate = {2023-08-09},
	author = {Hartmann, Jean and Boulet, Christian and Robert, Daniel},
	month = jan,
	year = {2021},
	doi = {10.1016/C2019-0-03488-7},
	publisher={Elsevier},
}

@article{niraula_impending_2022,
	title = {The impending opacity challenge in exoplanet atmospheric characterization},
	volume = {6},
	copyright = {2022 The Author(s), under exclusive licence to Springer Nature Limited},
	issn = {2397-3366},
	url = {https://www.nature.com/articles/s41550-022-01773-1},
	doi = {10.1038/s41550-022-01773-1},
	language = {en},
	number = {11},
	urldate = {2024-09-09},
	journal = {Nat. Astron},
	author = {Niraula, Prajwal and de Wit, Julien and Gordon, Iouli E. and Hargreaves, Robert J. and Sousa-Silva, Clara and Kochanov, Roman V.},
	month = nov,
	year = {2022},
	note = {Publisher: Nature Publishing Group},
	pages = {1287--1295},
}

@article{fortney2019need,
  title={The need for laboratory measurements and ab initio studies to aid understanding of exoplanetary atmospheres},
  author={Fortney, Jonathan J and Robinson, Tyler D and Domagal-Goldman, Shawn and Del Genio, Anthony D and Gordon, Iouli E and Gharib-Nezhad, Ehsan and Lewis, Nikole and Sousa-Silva, Clara and Airapetian, Vladimir and Drouin, Brian and others},
  journal={arXiv preprint arXiv:1905.07064},
  year={2019}
}

@article{ngo_isolated_2013,
	title = {An isolated line-shape model to go beyond the {Voigt} profile in spectroscopic databases and radiative transfer codes},
	volume = {129},
	issn = {0022-4073},
	url = {https://www.sciencedirect.com/science/article/pii/S0022407313002422},
	doi = {10.1016/j.jqsrt.2013.05.034},
	urldate = {2024-12-09},
	journal = {J. Quant. Spectrosc. Radiat. Transf.},
	author = {Ngo, N. H. and Lisak, D. and Tran, H. and Hartmann, J. -M.},
	month = nov,
	year = {2013},
	pages = {89--100},
}

@article{Gharib-Nezhad_2019,
	title = {The Influence of H2O Pressure Broadening in High-metallicity Exoplanet Atmospheres},
	volume = {872},
	issn = {},
	shorttitle = {},
	url = {https://doi.org/10.3847/1538-4357/aafb7b},
	doi = {10.3847/1538-4357/aafb7b},
	language = {en},
	number = {1},
	urldate = {},
	journal = {ApJ},
	author = {Gharib-Nezhad, Ehsan and Line, Michael R.},
	month = feb,
	year = {2019},
	pages = {27},
}

@article{Tan_JGR_2019,
author = {Tan, Y. and Kochanov, R. V. and Rothman, L. S. and Gordon, I. E.},
title = {Introduction of Water-Vapor Broadening Parameters and Their Temperature-Dependent Exponents Into the HITRAN Database: Part I—CO2, N2O, CO, CH4, O2, NH3, and H2S},
journal = {Geophys. Res. Atmos.},
volume = {124},
number = {21},
pages = {11580-11594},
doi = {https://doi.org/10.1029/2019JD030929},
url = {https://agupubs.onlinelibrary.wiley.com/doi/abs/10.1029/2019JD030929},
eprint = {https://agupubs.onlinelibrary.wiley.com/doi/pdf/10.1029/2019JD030929},
year = {2019}
}

@article{WILZEWSKI_HITRAN_2016,
title = {H2, He, and CO2 line-broadening coefficients, pressure shifts and temperature-dependence exponents for the HITRAN database. Part 1: SO2, NH3, HF, HCl, OCS and C2H2},
journal = {J. Quant. Spectrosc. Radiat. Transf.},
volume = {168},
pages = {193-206},
year = {2016},
issn = {0022-4073},
doi = {https://doi.org/10.1016/j.jqsrt.2015.09.003},
url = {https://www.sciencedirect.com/science/article/pii/S0022407315002988},
author = {Jonas S. Wilzewski and Iouli E. Gordon and Roman V. Kochanov and Christian Hill and Laurence S. Rothman}
}

@article{Wiesenfeld_2025,
doi = {10.3847/1538-4357/adb02e},
url = {https://doi.org/10.3847/1538-4357/adb02e},
year = {2025},
month = {mar},
publisher = {The American Astronomical Society},
volume = {981},
number = {2},
pages = {148},
author = {Wiesenfeld, Laurent and Niraula, Prajwal and de Wit, Julien and Jaïdane, Nejmeddine and Gordon, Iouli E. and Hargreaves, Robert J.},
title = {Ab Initio Quantum Dynamics as a Scalable Solution to the Exoplanet Opacity Challenge: A Case Study of CO2 in a Hydrogen Atmosphere},
journal = {ApJ}
}

@article{PADMANABHAN_JQSRT_2014,
title = {Study of pressure broadening effects of H2 on CO2 and CO in the near infrared region between 6317 and 6335cm−1 at room temperature},
journal = {J. Quant. Spectrosc. Radiat. Transf.},
volume = {133},
pages = {81-90},
year = {2014},
issn = {0022-4073},
doi = {https://doi.org/10.1016/j.jqsrt.2013.07.016},
url = {https://www.sciencedirect.com/science/article/pii/S0022407313002999},
author = {A. Padmanabhan and T. Tzanetakis and A. Chanda and M.J. Thomson}
}

@techreport{Hanson_Whitty_2014,
  author       = {Hanson, Ronald and Whitty, Kevin},
  title        = {Tunable Diode Laser Sensors to Monitor Temperature and Gas Composition in High-Temperature Coal Gasifiers},
  institution  = {Stanford Univ., CA (United States)},
  doi          = {10.2172/1222583},
  url          = {https://www.osti.gov/biblio/1222583},
  place        = {United States},
  year         = {2014},
  month        = {11}}

@article{Burch_JOSA_1969,
author = {Darrell E. Burch and David A. Gryvnak and Richard R. Patty and Charlotte E. Bartky},
journal = {J. Opt. Soc. Am.},
number = {3},
pages = {267--280},
publisher = {Optica Publishing Group},
title = {Absorption of Infrared Radiant Energy by CO2 and H2O. IV. Shapes of Collision-Broadened CO2 Lines$\ast$},
volume = {59},
month = {Mar},
year = {1969},
url = {https://opg.optica.org/abstract.cfm?URI=josa-59-3-267},
doi = {10.1364/JOSA.59.000267},
}

@book{Allen-Tidesley,
    author = {M. P. Allen and D. J. Tildesley},
    title = {Computer Simulation of Liquids},
    publisher = {Oxford: Oxford University Press},
    year = {1987},
}

@article{Hartmann2013a,
  title = {$Ab\phantom{\rule{0.28em}{0ex}}\phantom{\rule{0.28em}{0ex}}initio$ calculations of the spectral shapes of CO${}_{2}$ isolated lines including non-Voigt effects and comparisons with experiments},
  author = {Hartmann, J.-M. and Tran, H. and Ngo, N. H. and Landsheere, X. and Chelin, P. and Lu, Y. and Liu, A.-W. and Hu, S.-M. and Gianfrani, L. and Casa, G. and Castrillo, A. and Lep\`ere, M. and Deli\`ere, Q. and Dhyne, M. and Fissiaux, L.},
  journal = {Phys. Rev. A},
  volume = {87},
  issue = {1},
  pages = {013403},
  numpages = {11},
  year = {2013},
  month = {Jan},
  publisher = {American Physical Society},
  doi = {10.1103/PhysRevA.87.013403},
  url = {https://link.aps.org/doi/10.1103/PhysRevA.87.013403}
}

@article{Nguyen_CO2N2_JCP_2018,
    author = {Nguyen, H. T. and Ngo, N. H. and Tran, H.},
    title = {Prediction of line shape parameters and their temperature dependences for CO2–N2 using molecular dynamics simulations},
    journal = {J. Chem. Phys.},
    volume = {149},
    number = {22},
    pages = {224301},
    year = {2018},
    month = {12},
    issn = {0021-9606},
    doi = {10.1063/1.5063892},
    url = {https://doi.org/10.1063/1.5063892},
    eprint = {https://pubs.aip.org/aip/jcp/article-pdf/doi/10.1063/1.5063892/15551289/224301_1_online.pdf},
}

@article{TRAN_CMDS_O2_2019,
title = {Prediction of high-order line-shape parameters for air-broadened O2 lines using requantized classical molecular dynamics simulations and comparison with measurements},
journal = {J. Quant. Spectrosc. Radiat. Transf.},
volume = {222-223},
pages = {108-114},
year = {2019},
issn = {0022-4073},
doi = {https://doi.org/10.1016/j.jqsrt.2018.10.013},
url = {https://www.sciencedirect.com/science/article/pii/S0022407318306551},
author = {D.D. Tran and V.T. Sironneau and J.T. Hodges and R. Armante and J. Cuesta and H. Tran}
}

@article{NGUYEN_CMDS_CO2_2020,
title = {Line-shape parameters and their temperature dependences predicted from molecular dynamics simulations for O2- and air-broadened CO2 lines},
journal = {J. Quant. Spectrosc. Radiat. Transf.},
volume = {242},
pages = {106729},
year = {2020},
issn = {0022-4073},
doi = {https://doi.org/10.1016/j.jqsrt.2019.106729},
url = {https://www.sciencedirect.com/science/article/pii/S0022407319307320},
author = {H.T. Nguyen and N.H. Ngo and H. Tran}
}

@article{NGO_CMDS_N2O_2021,
title = {Air-broadened N2O line-shape parameters and their temperature dependences by requantized classical molecular dynamics simulations},
journal = {J. Quant. Spectrosc. Radiat. Transf.},
volume = {267},
pages = {107607},
year = {2021},
issn = {0022-4073},
doi = {https://doi.org/10.1016/j.jqsrt.2021.107607},
url = {https://www.sciencedirect.com/science/article/pii/S002240732100100X},
author = {N.H. Ngo and H.T. Nguyen and M.T. Le and H. Tran}
}

@article{HENDAOUI_CMDS_CO_2024,
title = {Refined line-shape parameters for CO lines broadened by air predicted from requantized classical molecular dynamics simulations},
journal = {J. Quant. Spectrosc. Radiat. Transf.},
volume = {319},
pages = {108954},
year = {2024},
issn = {0022-4073},
doi = {https://doi.org/10.1016/j.jqsrt.2024.108954},
url = {https://www.sciencedirect.com/science/article/pii/S002240732400061X},
author = {F. Hendaoui and H.T. Nguyen and H. Aroui and N.H. Ngo and H. Tran}
}

@article{NGO_Tran_CMDS_CO2_2025,
title = {Line-shape parameters and their temperature dependence for self-broadened CO2 lines in the 296 K- 1250 K range by requantized classical molecular dynamics simulations},
journal = {J. Quant. Spectrosc. Radiat. Transf.},
volume = {331},
pages = {109264},
year = {2025},
issn = {0022-4073},
doi = {https://doi.org/10.1016/j.jqsrt.2024.109264},
url = {https://www.sciencedirect.com/science/article/pii/S0022407324003716},
author = {N.H. Ngo and H. Tran}
}

@article{TRAN_CMDS_CO2N2_2025,
title = {Complete set of broadening coefficients and high-order line-shape parameters for N2-broadened CO2 lines, for temperatures ranging from 100 K to 1000 K},
journal = {J. Quant. Spectrosc. Radiat. Transf.},
volume = {342},
pages = {109499},
year = {2025},
issn = {0022-4073},
doi = {https://doi.org/10.1016/j.jqsrt.2025.109499},
url = {https://www.sciencedirect.com/science/article/pii/S002240732500161X},
author = {H. Tran and L. Denis and M. Lepère and B. Vispoel and N.H. Ngo}
}

@article{HENDAOUI_CMDS_CO2He_2026,
title = {A complete list of He-pressure-broadening coefficient of CO2 lines from 100 K to 3000 K for planet and exoplanet opacity calculations},
journal = {Icarus},
volume = {445},
pages = {116861},
year = {2026},
issn = {0019-1035},
doi = {https://doi.org/10.1016/j.icarus.2025.116861},
url = {https://www.sciencedirect.com/science/article/pii/S0019103525004099},
author = {Faten Hendaoui and Jean-Michel Hartmann and Hassen Aroui and Ha Tran}
}

@article{Hartmann_JCP_2010,
    author = {Hartmann, J.-M. and Boulet, C. and Tran, H. and Nguyen, M. T.},
    title = {Molecular dynamics simulations for CO2 absorption spectra. I. Line broadening and the far wing of the ν3 infrared band},
    journal = {J. Chem. Phys.},
    volume = {133},
    number = {14},
    pages = {144313},
    year = {2010},
    month = {10},
    issn = {0021-9606},
    doi = {10.1063/1.3489349},
    url = {https://doi.org/10.1063/1.3489349},
    eprint = {https://pubs.aip.org/aip/jcp/article-pdf/doi/10.1063/1.3489349/15432839/144313_1_online.pdf},
}

@article{Hartmann_CMDS_CO2H2O_2018,
    author = {Hartmann, Jean-Michel and Boulet, Christian and Tran, Duc Dung and Tran, Ha and Baranov, Yury},
    title = {Effect of humidity on the absorption continua of CO2 and N2 near 4 μm: Calculations, comparisons with measurements, and consequences for atmospheric spectra},
    journal = {J. Chem. Phys.},
    volume = {148},
    number = {5},
    pages = {054304},
    year = {2018},
    month = {02},
    issn = {0021-9606},
    doi = {10.1063/1.5019994},
    url = {https://doi.org/10.1063/1.5019994},
    eprint = {https://pubs.aip.org/aip/jcp/article-pdf/doi/10.1063/1.5019994/15537915/054304_1_online.pdf},
}

@article{Hartmann_CIA_CO2_2011,
    author = {Hartmann, J.-M. and Boulet, C.},
    title = {Molecular dynamics simulations for CO2 spectra. III. Permanent and collision-induced tensors contributions to light absorption and scattering},
    journal = {J. Chem. Phys.},
    volume = {134},
    number = {18},
    pages = {184312},
    year = {2011},
    month = {05},
    issn = {0021-9606},
    doi = {10.1063/1.3589143},
    url = {https://doi.org/10.1063/1.3589143},
    eprint = {https://pubs.aip.org/aip/jcp/article-pdf/doi/10.1063/1.3589143/14062707/184312_1_online.pdf},
}

@article{Hartmann_CMDS_CIA_N2_2017,
author = {Hartmann, J.-M. and Boulet, C. and Toon, G. C.},
title = {Collision-induced absorption by N2 near 2.16 µm: Calculations, model, and consequences for atmospheric remote sensing},
journal = {Geophys. Res. Atmos.},
volume = {122},
number = {4},
pages = {2419-2428},
doi = {https://doi.org/10.1002/2016JD025677},
url = {https://agupubs.onlinelibrary.wiley.com/doi/abs/10.1002/2016JD025677},
eprint = {https://agupubs.onlinelibrary.wiley.com/doi/pdf/10.1002/2016JD025677},
year = {2017}
}

@article{FAKHARDJI_CIA_CH2CO2_2022,
title = {Direct calculations of the CH4+CO2 far infrared collision-induced absorption},
journal = {J. Quant. Spectrosc. Radiat. Transf.},
volume = {283},
pages = {108148},
year = {2022},
issn = {0022-4073},
doi = {https://doi.org/10.1016/j.jqsrt.2022.108148},
url = {https://www.sciencedirect.com/science/article/pii/S0022407322000851},
author = {Fakhardji, Wissam and Boulet, Christian and Tran, Ha and Hartmann, Jean-Michel}
}

@article{HASHEMI_CO2_JQSRT_2020,
title = {Revising the line-shape parameters for air- and self-broadened CO2 lines toward a sub-percent accuracy level},
journal = {J. Quant. Spectrosc. Radiat. Transf.},
volume = {256},
pages = {107283},
year = {2020},
issn = {0022-4073},
doi = {https://doi.org/10.1016/j.jqsrt.2020.107283},
url = {https://www.sciencedirect.com/science/article/pii/S0022407320304751},
author = {Robab Hashemi and Iouli E. Gordon and Ha Tran and Roman V. Kochanov and Ekaterina V. Karlovets and Yan Tan and Julien Lamouroux and Ngoc Hoa Ngo and Laurence S. Rothman}
}

@article{Hellmann_CO2H2_PES_2025,
title = {Cross Second Virial Coefficients of the N2–H2, O2–H2, and CO2–H2 Systems from First Principles},
journal = {Int. J. Thermophys.},
volume = {46:67},
pages = {},
year = {2025},
issn = {1572-9567},
doi = {10.1007/s10765-025-03524-6},
url = {https://doi.org/10.1007/s10765-025-03524-6},
author = {Robert Hellmann  and  Eckard Bich}
}

@article{garland2019_arXiv,
  title={Effectively Calculating Gaseous Absorption in Radiative Transfer Models of Exoplanetary and Brown Dwarf Atmospheres},
  author={Garland, Ryan and Irwin, Patrick GJ},
  journal={arXiv preprint arXiv:1903.03997},
  year={2019}
}

@article{Goyal_ATMO_MNRAS_2017,
    author = {Goyal, Jayesh M and Mayne, Nathan and Sing, David K and Drummond, Benjamin and Tremblin, Pascal and Amundsen, David S and Evans, Thomas and Carter, Aarynn L and Spake, Jessica and Baraffe, Isabelle and Nikolov, Nikolay and Manners, James and Chabrier, Gilles and Hebrard, Eric},
    title = {A library of ATMO forward model transmission spectra for hot Jupiter exoplanets},
    journal = {MNRAS},
    volume = {474},
    number = {4},
    pages = {5158-5185},
    year = {2017},
    month = {11},
    issn = {0035-8711},
    doi = {10.1093/mnras/stx3015},
    url = {https://doi.org/10.1093/mnras/stx3015},
    eprint = {https://academic.oup.com/mnras/article-pdf/474/4/5158/23170529/stx3015.pdf},
}

@article{CO2_JW_Nature_2023,
    author = {Ahrer, Eva-Maria
and Alderson, Lili
and Batalha, Natalie M.
and Batalha, Natasha E.
and Bean, Jacob L.
and Beatty, Thomas G.
and Bell, Taylor J.
and Benneke, Björn
and Berta-Thompson, Zachory K.
and Carter, Aarynn L.
and Crossfield, Ian J. M.
and Espinoza, Néstor
and Feinstein, Adina D.
and Fortney, Jonathan J.
and Gibson, Neale P.
and Goyal, Jayesh M.
and Kempton, Eliza M.-R.
and Kirk, James
and Kreidberg, Laura
and López-Morales, Mercedes
and Line, Michael R.
and Lothringer, Joshua D.
and Moran, Sarah E.
and Mukherjee, Sagnick
and Ohno, Kazumasa
and Parmentier, Vivien
and Piaulet, Caroline
and Rustamkulov, Zafar
and Schlawin, Everett
and Sing, David K.
and Stevenson, Kevin B.
and Wakeford, Hannah R.
and Allen, Natalie H.
and Birkmann, Stephan M.
and Brande, Jonathan
and Crouzet, Nicolas
and Cubillos, Patricio E.
and Damiano, Mario
and Désert, Jean-Michel
and Gao, Peter
and Harrington, Joseph
and Hu, Renyu
and Kendrew, Sarah
and Knutson, Heather A.
and Lagage, Pierre-Olivier
and Leconte, Jérémy
and Lendl, Monika
and MacDonald, Ryan J.
and May, E. M.
and Miguel, Yamila
and Molaverdikhani, Karan
and Moses, Julianne I.
and Murray, Catriona Anne
and Nehring, Molly
and Nikolov, Nikolay K.
and Petit dit de la Roche, D. J. M.
and Radica, Michael
and Roy, Pierre-Alexis
and Stassun, Keivan G.
and Taylor, Jake
and Waalkes, William C.
and Wachiraphan, Patcharapol
and Welbanks, Luis
and Wheatley, Peter J.
and Aggarwal, Keshav
and Alam, Munazza K.
and Banerjee, Agnibha
and Barstow, Joanna K.
and Blecic, Jasmina
and Casewell, S. L.
and Changeat, Quentin
and Chubb, K. L.
and Colón, Knicole D.
and Coulombe, Louis-Philippe
and Daylan, Tansu
and de Val-Borro, Miguel
and Decin, Leen
and Dos Santos, Leonardo A.
and Flagg, Laura
and France, Kevin
and Fu, Guangwei
and García Muñoz, A.
and Gizis, John E.
and Glidden, Ana
and Grant, David
and Heng, Kevin
and Henning, Thomas
and Hong, Yu-Cian
and Inglis, Julie
and Iro, Nicolas
and Kataria, Tiffany
and Komacek, Thaddeus D.
and Krick, Jessica E.
and Lee, Elspeth K. H.
and Lewis, Nikole K.
and Lillo-Box, Jorge
and Lustig-Yaeger, Jacob
and Mancini, Luigi
and Mandell, Avi M.
and Mansfield, Megan
and Marley, Mark S.
and Mikal-Evans, Thomas
and Morello, Giuseppe
and Nixon, Matthew C.
and Ortiz Ceballos, Kevin
and Piette, Anjali A. A.
and Powell, Diana
and Rackham, Benjamin V.
and Ramos-Rosado, Lakeisha
and Rauscher, Emily
and Redfield, Seth
and Rogers, Laura K.
and Roman, Michael T.
and Roudier, Gael M.
and Scarsdale, Nicholas
and Shkolnik, Evgenya L.
and Southworth, John
and Spake, Jessica J.
and Steinrueck, Maria E.
and Tan, Xianyu
and Teske, Johanna K.
and Tremblin, Pascal
and Tsai, Shang-Min
and Tucker, Gregory S.
and Turner, Jake D.
and Valenti, Jeff A.
and Venot, Olivia
and Waldmann, Ingo P.
and Wallack, Nicole L.
and Zhang, Xi
and Zieba, Sebastian
and JWST Transiting Exoplanet Community Early Release Science Team
},
    title = {Identification of carbon dioxide in an exoplanet atmosphere},
    journal = {Nature},
    volume = {614},
    number = {},
    pages = {649-652},
    year = {2023},
    month = {02},
    issn = {7949},
    doi = {10.1038/s41586-022-05269-w},
    url = {https://doi.org/10.1038/s41586-022-05269-w},
    eprint = {},
}

@article{Tremblin_2015,
doi = {10.1088/2041-8205/804/1/L17},
url = {https://doi.org/10.1088/2041-8205/804/1/L17},
year = {2015},
month = {apr},
publisher = {The American Astronomical Society},
volume = {804},
number = {1},
pages = {L17},
author = {Tremblin, P. and Amundsen, D. S. and Mourier, P. and Baraffe, I. and Chabrier, G. and Drummond, B. and Homeier, D. and Venot, O.},
title = {FINGERING CONVECTION AND CLOUDLESS MODELS FOR COOL BROWN DWARF ATMOSPHERES},
journal = {ApJL}
}

@article{Balmer_2025,
doi = {10.3847/1538-3881/adb1c6},
url = {https://doi.org/10.3847/1538-3881/adb1c6},
year = {2025},
month = {mar},
publisher = {The American Astronomical Society},
volume = {169},
number = {4},
pages = {209},
author = {Balmer, William O. and Kammerer, Jens and Pueyo, Laurent and Perrin, Marshall D. and Girard, Julien H. and Leisenring, Jarron M. and Lawson, Kellen and Dennen, Henry and van der Marel, Roeland P. and Beichman, Charles A. and Bryden, Geoffrey and Llop-Sayson, Jorge and Valenti, Jeff A. and Lothringer, Joshua D. and Lewis, Nikole K. and Mâlin, Mathilde and Rebollido, Isabel and Rickman, Emily and Hoch, Kielan K. W. and Soummer, Rémi and Clampin, Mark and Mountain, C. Matt},
title = {JWST-TST High Contrast: Living on the Wedge, or, NIRCam Bar Coronagraphy Reveals CO2 in the HR 8799 and 51 Eri Exoplanets’ Atmospheres},
journal = {Astron. J.}
}

@article{Hase:99,
author = {Frank Hase and Thomas Blumenstock and Clare Paton-Walsh},
journal = {Appl. Opt.},
number = {15},
pages = {3417--3422},
publisher = {Optica Publishing Group},
title = {Analysis of the instrumental line shape of high-resolution Fourier transform IR spectrometers with gas cell measurements and new retrieval software},
volume = {38},
month = {May},
year = {1999},
url = {https://opg.optica.org/ao/abstract.cfm?URI=ao-38-15-3417},
doi = {10.1364/AO.38.003417},
}

@Article{Hase_amt-5-603-2012,
AUTHOR = {Hase, F.},
TITLE = {Improved instrumental line shape monitoring for the ground-based, high-resolution FTIR spectrometers of the Network for the Detection of Atmospheric Composition Change},
JOURNAL = {Atmos. Meas. Tech.},
VOLUME = {5},
YEAR = {2012},
NUMBER = {3},
PAGES = {603--610},
URL = {https://amt.copernicus.org/articles/5/603/2012/},
DOI = {10.5194/amt-5-603-2012}
}

@ARTICLE{Rosenkranz_1975,

  author={Rosenkranz, P.},

  journal={IEEE Trans. Antennas Propag.}, 

  title={Shape of the 5 mm oxygen band in the atmosphere}, 

  year={1975},

  volume={23},

  number={4},

  pages={498-506},

  doi={10.1109/TAP.1975.1141119}}

@article{nelkin1964simple,
  title={Simple binary collision model for Van Hove's G s (r, t)},
  author={Nelkin, Mark and Ghatak, Ajoy},
  journal={Phys. Rev.},
  volume={135},
  number={1A},
  pages={A4},
  year={1964},
  publisher={APS}
}

@article{NIE_2025,
title = {Towards improved spectroscopic applications in multi-gas environments: CO2 spectral line parameter measurement and analysis},
journal = {Spectrochim. Acta A Mol. Biomol. Spectrosc.},
volume = {328},
pages = {125428},
year = {2025},
issn = {1386-1425},
doi = {https://doi.org/10.1016/j.saa.2024.125428},
url = {https://www.sciencedirect.com/science/article/pii/S1386142524015944},
author = {Wei Nie and Zhongzheng Zhou and Zhenyu Xu and Rantong Niu and Yuzhou Ran and Ruifeng Kan}
}

@article{mHT_2025,
title = {New beyond-Voigt line-shape profile recommended for the HITRAN database},
journal = {J. Quant. Spectrosc. Radiat. Transf.},
volume = {347},
pages = {109596},
year = {2025},
issn = {0022-4073},
doi = {https://doi.org/10.1016/j.jqsrt.2025.109596},
url = {https://www.sciencedirect.com/science/article/pii/S0022407325002584},
author = {P. Wcisło and N. Stolarczyk and M. Słowiński and H. Jóźwiak and D. Lisak and R. Ciuryło and A. Cygan and F. Schreier and C.D. Boone and A. Castrillo and L. Gianfrani and Y. Tan and S.-M. Hu and E.M. Adkins and J.T. Hodges and H. Tran and H.N. Ngo and J.-M. Hartmann and S. Beguier and A. Campargue and R.J. Hargreaves and L.S. Rothman and I.E. Gordon}
}

@article{MONDELAIN2025109271,
title = {Isotopologue dependence of the CO2-air broadening and shifting coefficients: Experimental evidence and comparison with theory for 13CO2 and 12CO2},
journal = {J. Quant. Spectrosc. Radiat. Transf.},
volume = {333},
pages = {109271},
year = {2025},
issn = {0022-4073},
doi = {https://doi.org/10.1016/j.jqsrt.2024.109271},
url = {https://www.sciencedirect.com/science/article/pii/S0022407324003789},
author = {Didier Mondelain and Alain Campargue and Robert R. Gamache and Jean-Michel Hartmann and Fabien Gibert and Georg Wagner and Manfred Birk and Christian Röske}
}

@article{hedges2016effect,
  title={Effect of pressure broadening on molecular absorption cross sections in exoplanetary atmospheres},
  author={Hedges, Christina and Madhusudhan, Nikku},
  journal={MNRAS},
  volume={458},
  number={2},
  pages={1427--1449},
  year={2016},
  publisher={The Royal Astronomical Society}
}

@article{gamache2018temperature,
  title={On the temperature dependence of half-widths and line shifts for molecular transitions in the microwave and infrared regions},
  author={Gamache, Robert R and Vispoel, Bastien},
  journal={J. Quant. Spectrosc. Radiat. Transf.},
  volume={217},
  pages={440--452},
  year={2018},
  publisher={Elsevier}
}

\clearpage
\onecolumn

\appendix
\section{Additional table}
\label{appendix}

\begin{table}[h]
	  \caption{Sample of parameters obtained from fits of the single power and double power laws to rCMDS results. }
      \centering
   	\begin{tabular}{r c c c c c c c }
        \hline  \hline 
        $m$ & $\gamma_{\mathrm{H_2}}$ &  $n$ &    & $\gamma_{0}$ & $n_1$ &  $\gamma'_{0}$ & $n_2$\\
        \hline 
       
        1 & 0.1382 (12)  &  0.761 (19) &   & 0.1225 (25) & 0.641 (112) & 0.0130 (229) & 2.204 (1549) \\
        2 & 0.1280 (6) & 0.728 (10) &   & 0.1248 (28) & 0.782 (22) & 0.0024 (27) & -0.785 (615) \\
        3 & 0.1221 (7) & 0.719 (11) &   & 0.1178 (7) & 0.786 (5) & 0.0034 (7) & -0.641 (108) \\ 
        4 & 0.1191 (7) & 0.713 (12) &   & 0.1122 (12) & 0.801 (8) & 0.0059 (12) & -0.383 (96)\\
        5 & 0.1173 (7) & 0.709 (12) &   & 0.1063 (23) & 0.822 (13) & 0.0098 (23) & -0.169 (95) \\
        6 & 0.1160 (7) & 0.706 (12) &   & 0.1009 (35) & 0.843 (18) & 0.0140 (35) & -0.037 (93) \\
        7 & 0.1152 (7) & 0.704 (12) &   & 0.0961 (41) & 0.861 (20) & 0.0179 (41) & 0.050 (80)\\
        8 & 0.1145 (7) & 0.702 (12) &   & 0.0912 (53) & 0.881 (25) & 0.0221 (53) & 0.120 (79)\\
        9 & 0.1140 (7) & 0.701 (12)&   & 0.0862 (66) & 0.902 (31) & 0.0266 (66) & 0.178 (78)\\
        10 &0.1136 (7)& 0.700 (13) &   &0.0832 (67)& 0.914 (31) &0.0292 (66) & 0.205 (71)\\
        11&0.1133 (7)& 0.699 (13)&   &0.0790 (75)&0.933 (36)& 0.0331 (75)& 0.243 (69)\\
        12& 0.1130 (7) & 0.699 (13)&   &0.0770 (72)& 0.942 (35) & 0.0347 (72) & 0.257 (63)\\
        13&0.1127 (7)& 0.698 (13)&   &0.0744 (69)& 0.954 (35)& 0.0371 (69) & 0.276 (56)\\
        14&0.1125 (7) &	0.698 (13)&   &0.0734 (60)&0.958 (30)& 0.0379 (60)&	0.283 (47)\\	
        15&0.1124 (7)&0.698(13)&   &0.0718(58) &0.966 (30)&	0.0393 (58)&0.294 (44)\\
        16&0.1122 (7)&	0.697 (13)&   &0.0697 (54)& 0.976 (29)&	0.0413 (54)&0.308 (39)\\
        17&	0.1121 (7)&	0.697 (13)&	  &0.0680 (45)&0.985 (24)&	0.0429 (45)&	0.319 (31)\\
        18&	0.1120 (7)& 0.697 (13)&   &0.0678 (26)&0.986 (14)&	0.0429 (26)&	0.319 (18)\\
        19&	0.1118 (7)&	0.697 (13)&   &0.0676 (27)&0.987 (14)&	0.0429 (27)&	0.319 (18)\\
        20&	0.1117 (7)&	0.697 (13)&   &0.0676 (23)&0.986 (13)&	0.0429 (23)&	0.319 (16)\\
        21&	0.1117 (7)&	0.697 (13)&   &0.0667 (28)&0.991 (15)&  0.0437 (28)&	0.325 (19)\\
        22&	0.1116 (7)&	0.697 (13)&   &0.0681 (42)&0.984 (23)& 	0.0422 (42)&	0.314 (30)\\
        23& 0.1115 (7)&	0.697 (13)&   &0.0667 (49)&0.990 (27)&  0.0435 (49)&	0.324 (33)\\
        24&	0.1114 (7)& 0.697 (13)&   &0.0667 (66)&0.991 (36)&	0.0435 (66)&	0.323 (45)\\
        25&	0.1114 (7)&	0.697 (13)&   &0.0672 (78)&0.988 (42)&	0.0429 (77)&	0.319 (54)\\
        26&	0.1113 (7)&	0.697 (13)&   &0.0673 (89)&0.987 (48)&	0.0427 (88)&	0.318 (62)\\
        27&	0.1112 (7)& 0.697 (13)&   & 0.0679 (99)&0.984 (54)&  0.0420 (99)&	0.313 (70)\\
        28&	0.1112 (7)&	0.697 (13)&   &	0.0682 (106)&0.982 (57)& 0.0417 (106)&	0.311 (76)\\
        29&	0.1111 (7)&	0.697 (13)&   & 0.0685 (125)&0.981 (67)& 0.0414 (124)&	0.309 (90)\\
        30&	0.1111 (7)&	0.697 (13)&   &0.0698 (138)&0.974 (73)&	0.0400 (138)& 0.299 (103)\\
        31& 0.1110 (7)& 0.697 (13)&   &0.0701 (143)&0.972 (76)& 0.0397 (143)& 0.296 (108)\\
        32&	0.1110 (7)&	0.697 (13)&   &0.0706 (168)&0.970 (89)& 0.0391 (168)& 0.292 (129)\\
        33&0.1110 (7)& 0.697 (13)&   &0.0703 (178)&0.972 (94)& 0.0394 (177)& 0.295 (135)\\
        34&0.1109 (7)&0.697 (13)&   &0.0716 (187)&0.964 (97)& 0.0380 (186)& 0.284 (148)\\
        35&0.1109 (7)&0.697 (13)&   & 0.0715 (203) &0.965 (106) &0.0381 (203)&	0.285 (160)\\
        36&0.1109 (7)&0.698 (13)&   & 0.0725 (211) &0.960 (109) &0.0371 (210)& 0.277 (172)\\
        37&0.1108 (7)&	0.698 (13)&   &0.0734 (230)&0.955 (119) &0.0362 (230)&	0.269 (193)\\
        38&	0.1108 (7)&	0.698 (13)&   &0.0728 (237)&0.959 (123) &0.0367 (236)&	0.274 (195)\\
        39&	0.1108 (7)&	0.698 (13)&   &	0.0748 (246)&0.949 (126)&0.0348 (246)&	0.257 (216)\\
        40&	0.1107 (7)&	0.698 (13)&   &0.0752 (255)&	0.946 (130)&	0.0343 (255)&0.253 (227)\\
        41&	0.1107 (7)&	0.698 (13)&   &0.0757 (273)&	0.944 (138)&	0.0338 (272)&0.249 (247)\\
        42& 0.1107 (7)&	0.698 (13)&   &0.0761 (282)&	0.942 (142)&	0.0333 (281)&0.245 (259)\\
        43& 0.1107 (7)& 0.698 (13)&   & 0.0765 (290)&   0.940 (146)& 0.0329 (289)&    0.241 (271)\\
        44&	0.1106 (7)&	0.699 (13)&   &	0.0765 (299)&	0.940 (151)& 0.0328 (298)&	0.241 (280)\\
        45&	0.1106 (7)&	0.699 (13)&   &	0.0775 (312)&	0.935 (156)& 0.0319 (311)&	0.232 (302)\\
        46&	0.1106 (7)&	0.699 (13)&   &	0.0783 (311)&	0.931 (156)& 0.0310 (310)&	0.223 (312)\\
        47&	0.1106 (7)&	0.699 (13)&   & 0.0796 (318)&	0.925 (158)& 0.0297 (317)&	0.211 (334)\\
        48&	0.1105 (7)&	0.699 (13)&   &	0.0796 (326)&	0.925 (162)& 0.0297 (325)&	0.210 (344)\\
        49&	0.1105 (7)&	0.699 (13)&   &	0.0809 (328)&   0.919 (162)& 0.0284 (327)&	0.196 (364)\\
        50&	0.1105 (7)&	0.699 (13)&   &	0.0809 (346)&	0.919 (171)& 0.0284 (345)& 	0.196 (385)\\
        51&	0.1105 (7)& 0.700 (13)&   &	0.0817 (351)&	0.914 (174)& 0.0275	(350)&	0.187 (406)\\
        52&	0.1105 (7)&	0.700 (13)&   & 0.0821 (354)&	0.913 (175)& 0.0271 (352)&	0.181 (416)\\	

        \hline 
	\end{tabular}
    \tablefoot{%
$\gamma_{\mathrm{H_2}}$, $\gamma_{0}$ and $\gamma'_{0}$ are expressed in cm$^{-1}$/atm, while $n$, $n$$_1$ and $n$$_2$ are dimensionless.}

\end{table}

\end{document}